\documentclass{aa}
\usepackage[varg]{txfonts}
\usepackage{newtxtext,newtxmath}
\usepackage[T1]{fontenc}
\usepackage{xspace}
\usepackage{graphicx}
\usepackage{amsmath}
\usepackage[colorlinks=true,linkcolor=blue,citecolor=blue,urlcolor=blue]{hyperref}
\usepackage[linesnumbered,ruled]{algorithm2e}
\usepackage{booktabs}
\usepackage{pifont}

\newcommand{\RN}[1]{\textup{\uppercase\expandafter{\romannumeral#1}}}

\newcommand{\Lya}{\ifmmode{\mathrm{Ly}\alpha}\else Ly$\alpha$\xspace\fi}
\newcommand{\MgII}{Mg\RN{2}\xspace}
\newcommand{\FeII}{Fe\RN{2}\xspace}
\newcommand{\SiII}{Si\RN{2}\xspace}
\newcommand{\CIV}{C\RN{4}\xspace}
\newcommand{\thor}{\textsc{thor}\xspace}
\newcommand{\msun}{\,{M$_{\odot}$}\xspace}
\newcommand{\angstrom}{\text{\normalfont\AA}\xspace}
\newcommand{\sbunits}{\,erg\,s$^{-1}$\,cm$^{-2}$\,arcsec$^{-2}$\xspace}
\newcommand{\cmark}{\ding{51}}  \newcommand{\xmark}{\ding{55}}  

\makeatletter
\let\linenumbers\relax

\makeatother

\begin{document}

\title{THOR: a GPU-accelerated and MPI-parallel radiative transfer code}

\author{Chris Byrohl\inst{1,2}\thanks{E-mail: chris.byrohl@uni-heidelberg.de}
  \and Dylan Nelson\inst{1,3}} 

\institute{
Universität Heidelberg, Institut für Theoretische Astrophysik, ZAH, 
Albert-Ueberle-Str. 2, 69120 Heidelberg,
Germany
\and
Kavli IPMU (WPI), UTIAS, The University of Tokyo, Kashiwa, Chiba 277-8583, Japan
\and
Universität Heidelberg, Interdisziplinäres Zentrum für Wissenschaftliches Rechnen, INF 205, 69120 Heidelberg, Germany}

\date{}

\abstract{
Emission and absorption line features are important diagnostics for the physics underlying extragalactic astronomy. The interpretation of observed signatures involves comparing against forward modeled spectra from galaxy formation simulations as well as more simplified geometries, while including the complex scattering radiative transfer (RT) of resonant emission lines. Here, we present \thor, a modern C++ radiative transfer code focused initially on resonant emission lines. \thor is a high-performance, distributed memory MPI-parallel, multi-target code, running on CPUs, GPUs and other accelerators, yielding large $\sim 10-50\rm{x}$ speed-ups compared to previous CPU-only codes. We support multiple grid-based and gridless data structures, enabling comparisons across different hydrodynamical codes as well as toy model geometries. We demonstrate its science capabilities with a number of example use cases across scales: (i) Lyman-alpha RT on simple shell-like gas distributions; (ii) Lyman-alpha RT applied to a high-resolution, high-redshift $z \sim 6$ cosmological hydrodynamical galaxy formation simulation; (iii) Lyman-alpha and Magnesium-II halos, i.e.\ scattering and emission from the circumgalactic medium of $z \simeq 1-2$ galaxies drawn from cosmological magnetohydrodynamical simulations; (iv) the large-scale cosmic web in gas emission, a $6144^3$ volume element RT scaling calculation; and (v) synthetic absorption spectra of the Lyman-alpha forest. Extensive verification and benchmarking validates our approach and its computational efficiency.}


\maketitle

\section{Introduction}

Astrophysical observations of galaxies and their diffuse surroundings rely on detecting the light i.e. electromagnetic radiation that they emit. The interpretation of these observations allows us to study the distribution and physical properties of matter across its many phases. Fundamentally, radiation is our empirical window into the large-scale structure of the Universe, as well as the physics underlying galaxy formation and evolution. In order to gain insight from these observables -- from emission to absorption -- we must understand not only how radiation is emitted, but also how it propagates and interacts with astrophysical media across a variety of scales. The problem of radiative transfer (RT) is a common theme, from stars, to star-forming clouds, the interstellar medium (ISM) of galaxies, the circumgalactic medium (CGM) of dark matter halos, and the intergalactic medium (IGM) i.e. the large-scale cosmic web itself. 

\subsection{Science and observables of radiative transfer}

One powerful way to interpret such observations is by forward modeling the observational radiative signatures of theoretical models. These can take the form of simple, analytical configurations with one-dimensional geometry \citep[e.g][]{Neufeld90}, all the way to multi-scale, three-dimensional, full cosmological hydrodynamical simulations \citep[for a review see][]{vogelsberger2020}. In either case, solving the high dimensional, non-local equations of radiative transfer is challenging, and requires efficient and specialized numerical tools \citep{noebauer19}. This is particularly true for problems involving scattering, optically thick media, and resonant line emission, all of which are computationally demanding \citep{madau95}. In particular, an important gas observable across extragalactic astrophysics is the emission that arises from the radiative transitions of atoms, ions, and molecules \citep{osterbrock89}. As one of the brightest emission lines in the Universe, the \Lya transition of hydrogen is a prominent example \citep[for a recent review see][]{ouchi20}. 

\Lya is also an example of resonant line emission, that occurs for transitions between the ground state and first excited electronic state. If optical depths are high and other interactions, such as non-resonant lines or absorption by dust at nearby wavelengths are sufficiently small, photons can experience multiple scatterings by absorption and re-emission via the same transition. These scatterings qualitatively change spectral and spatial signatures. While this complicates the interpretation of resonant emission line sources, it simultaneously offers the opportunity to trace otherwise dark, diffuse media.

Resonance lines occur for neutral hydrogen as well as various ionization levels of metals, including Carbon, Nitrogen, Sodium, Magnesium, Oxygen, and Iron. These lines commonly fall in the rest-frame ultraviolet, and are readily observable from either ground and space, particularly when redshifted. Many of these lines are key tracers of the CGM and IGM, and examples include CIV~\citep[$\lambda\lambda 1448,1550$;][]{Cowie95, Lehner16}, \MgII~\citep[$\lambda\lambda 2796, 2803$;][]{Weymann79,Anand21}, and OVI~\citep[$\lambda\lambda 1032,1038$;][]{Danforth05, Tripp08}. At higher ionization levels, resonant emission lines also reach X-ray energies, such as Fe XVII at 15.01\AA\xspace and 17.05\AA\xspace, probing the halos of high-mass groups and clusters as well as the warm-hot intergalactic medium \citep{churazov10}.

When spatially resolved, emission observations provide powerful views on the abundance, physical state, and kinematics of gaseous halos. Spatially extended CGM emission has been detected in several lines, including \MgII~\citep{Rubin11,Zabl21,Leclercq22,Burchett21,Pessa24}, \SiII~\citep{Kusakabe24}, \FeII~\citep{Finley17}, and \CIV~\citep{Fossati21}. Observations of extended emission are particularly abundant for \Lya, as observed around quasars \citep[e.g.][]{heckman91,borisova16,arrigoni19} as well as star-forming galaxies~\citep[e.g.][]{Hayashino04,Steidel11,wisotzki16,Momose16,LujanNiemeyer22a}, where emission is detected up to megaparsec scales~\citep{Kakuma21,Kikuchihara22}. \Lya also traces nodes and even filaments of the large-scale cosmic web~\citep{Cantalupo14,Umehata19,Bacon21,Martin23}. Modern integral field spectrographs such as VLT-MUSE, KCWI, and HET-VIRUS~\citep{Bacon10,Morrissey18,Hill21} can concurrently map spatial and spectral signatures of the diffuse gas, identifying individual LAEs~\citep{Leclercq17} while also probing its volume filling spectral signatures~\citep{Leclercq20,Wang21}. New surveys and missions such as SPHEREx and ODIN will further add to the existing wealth of extended \Lya observations~\citep{Dore14,Lee24}.

The \Lya line and its absorption features, the \Lya forest, traces the neutral hydrogen distribution to the end of the Epoch of Reionization \citep[EoR, $z\sim 6$;][]{Gunn65,Mason18,Prochaska19}. It is an ubiquitous emission feature across spatial scales, from stellar atmosphere and nebular emission in the ISM to the cosmic web. It also occurs across redshift, from nearby galaxies at $z \sim 0$ into the EoR, giving important insights into stellar and galaxy evolution processes \citep{charlot93}, the large-scale matter distribution and our cosmological concordance model \citep{sargent80,hernquist96}, as well as the progression and sources of reionization \citep{miralda90,stark10,bhagwat25}. Notably, \Lya is a key observable of EoR galaxies and their surroundings, as now being probed by the James Webb Space Telescope \citep{witstok25,stark25}. \Lya may even be dynamically important for galaxy evolution, where modeling these effects requires coupled, on-the-fly calculations~\citep{Kimm18,Tomaselli21,Nebrin25}. Its various emission channels, high optical depths, and complex scattering behavior requires detailed forward-modeling \citep{camps21}. This makes modeling \Lya a difficult litmus test when modeling resonant emission lines, and thus it is a key focus of this paper.

\subsection{Numerical radiative transfer methods}

Several public, open-source codes for resonant emission line radiative transfer are available to the community, including LaRT, COLT and RASCAS~\citep{Seon20, Smith15, Michel-Dansac20}. These are based on the Monte Carlo radiative transfer (MCRT) technique, as also used by many other codes with a diversity of capabilities~\citep{Verhamme06, Behrens19, Byrohl21}.

There are also problem-specific MCRT codes, for ionizing RT e.g. \textsc{Mocassin} and \textsc{CMacIonize}~\citep{ercolano03,Vandenbroucke18}, dust RT e.g. \textsc{Radmc-3d}, \textsc{Hyperion} and \textsc{Skirt}~\citep{Dullemond12,robitaille11,camps15}, and supernovae post-processing e.g. \textsc{Sedona} and \textsc{Supernu}~\citep{Kasen06,Wollaeger13}. 

All astrophysical MCRT codes described above -- for resonant scattering, ionizing RT, and so on -- run on CPUs only, and none has support for modern hetereogeneous computing architectures based on graphical processing units (GPU). Some notable, domain-specific exceptions exist \citep[e.g.][]{Heymann12,Lee22,hirling24,matthews25}. These typically use the vendor-specific CUDA library API to enable GPU acceleration on Nvidia hardware. In addition, existing methods using the \textsc{OpenMP} framework for CPU parallelization can add high-level directives to enable GPU offloading~\citep[see e.g. the \textsc{ARTIS} code;][]{sim07}. However, this prevents use of important device-level concepts such as memory hierarchy and queued/asynchronous execution. Overall, GPU-accelerated astrophysical MCRT codes are still in their infancy, and more work is needed, with emission line MCRT being a clear use case.

Two major applications of emission line MCRT codes are (i) post-processing of hydrodynamical simulations to forward model observational signatures arising from simulated galaxies and their surroundings, and (ii) calculating the emergent radiation from simplified geometries, often for large numbers of model configuration e.g. to fit a particular observation. In both cases, computational demand can be substantial. As a result, substantial benefits -- and new scientific applications -- result from increasing the computational efficiency of these codes. The use of new hardware architectures plays a pivotal role in reaching this goal. Accelerators such as graphics processing units (GPUs) now typically provide the majority of the computational throughput of the largest supercomputers in the world. As a result, codes must increasingly support a diverse set of accelerators from different vendors, with AMD and Intel now common in addition to Nvidia.

In this paper, we introduce a new general purpose, high-performance, multi-device, distributed resonant emission line Monte Carlo radiative transfer code, \thor. We implement CPU and GPU support, and investigate its performance across diverse applications and system architectures. Combining the SYCL abstraction layer for broad GPU support, with inter-node communication via the MPI framework, \thor is an efficient, hybrid MPI+SYCL, exascale-ready code. We focus on resonant emission line radiative transfer as a first key application, touching briefly upon other use cases including generic ray-tracing, absorption sightlines, and volume rendering.

The paper is structured as follows. In Section~\ref{sec:rtphysics} we briefly describe the relevant radiative transfer physics. Section~\ref{sec:thor} then gives an overview of the \thor code, its architecture, and its numerical methodology. We validate its results on test problems in Section~\ref{sec:validation}, and then showcase a variety of scientific applications in Section~\ref{sec:applications}. In Section~\ref{sec:performance} we present performance benchmarks and scaling tests. Finally, Section~\ref{sec:discussion} discusses limitations and future directions, while Section~\ref{sec:summary} summarizes the work.

\section{Radiative Transfer -- Physical Concepts}
\label{sec:rtphysics}

In this section, we summarize the physics underlying resonant emission line radiative transfer and its numerical implementation. When we need to consider a specific transition, we adopt \Lya as a fiducial case; the same concepts and details apply for other resonant emission lines. We express the relevant physics in terms of individual photons, consistent with the design of our MCRT code that traces the evolution of radiation packets subject to the physical processes experienced by individual photons.

\subsection{Emission}

We conceptually separate the emission mechanism, i.e. the process leading to the emission of a photon, from its physical origins, i.e. the source providing the energy~\citep{Dijkstra17,byrohl22}.

\subsubsection{Mechanisms}

There are a range of emission channels responsible for the energy release of a given emission line. For \Lya, the two major mechanisms are collisional excitations of neutral hydrogen, as well as recombinations of ionized hydrogen:
\begin{align} \label{eq:lumrec}
\epsilon_\mathrm{rec} &= f_\mathrm{rec}(T)\,\alpha(T) \,n_\mathrm{e} \,n_\mathrm{HII} \,E_{\mathrm{Ly}\alpha}, \\
\epsilon_\mathrm{coll} &= \gamma_{\mathrm{1s2p}}(T) \,n_\mathrm{e} \,n_\mathrm{HI} \,E_{\mathrm{Ly}\alpha},
\end{align}
which scale with the number density of electrons ($n_\mathrm{e}$), neutral ($n_\mathrm{HI}$) and ionized hydrogen ($n_\mathrm{HII}$). The rates are described by the temperature-dependent recombination and collisional excitation coefficients $\alpha(T)$ and $\gamma_\mathrm{1s2p}(T)$. The fraction $f_\mathrm{rec}(T)$ gives the probability for \Lya emission upon recombination. We adopt case-B recombinations, i.e.\ under the assumption that Lyman-series photons are optically thick~\citep[taking coefficients from][]{Scholz90,Scholz91,Draine11}.

Emitted photons follow a given angular and spectral probability distribution in the rest frame of the emitting atom. We typically adopt isotropic emission that follows a thermal Gaussian or Lorentzian profile for the spectral shape, based on the particular emission line and atomic velocity distribution.

Other common photon emission mechanisms, such as free-free emission, can indirectly contribute to \Lya luminosities when subsequently scattered by the \Lya transition. This can either occur due to small continuum contributions near the line resonance, as well as from the Hubble-redshifted continuum between $916-1216$\angstrom~\citep{Silva13,Pullen14}.

\subsubsection{Origins}

For \Lya, most photons are ultimately emitted by collisional excitations and recombinations. The temperature, density and ionization state of hydrogen carries imprints from a number of important \Lya origins, such as gravitational cooling and shock heating, local and non-local ionizing radiation photoionization and photo-heating. This is often self-consistently captured in modern (cosmological) galaxy formation simulations.

Depending on the line, other origins are important. Emission can arise from stars themselves, local ionized regions of gas around bright young stars (i.e. HII regions), and active galactic nuclei (AGN).\footnote{The metagalactic background radiative field (UVB) is physically captured as inducing the emission channels discussed above.} In all cases, the physical emission model must be considered within the context of the numerics and resolution of a particular hydrodynamical simulation.

For the emission from stars, we can consider three classes of models: (i) the radiation emitted by individual stars, as appropriate for high-resolution, single-star type galaxy simulations; (ii) the radiation emitted by stellar populations, e.g. for large-volume simulations such as IllustrisTNG where star particles represent co-eval populations;\footnote{In this case, if the line emission of interest originates not from stars themselves, but from the reprocessing of stellar radiation in the surrounding ISM gas, a further modeling step is required, particularly when this gas is unresolved in the hydrodynamical simulation~\citep[e.g.][]{byler18,Byrohl23,kapoor24}.} or (iii) the radiation emitted by entire galaxies, e.g. for lower resolution simulations where galaxies are not internally resolved, and simple theoretical relations or empirical scalings with mass or star formation rate can set the total luminosity of the relevant emission line. 

\subsection{Propagation}
\label{sec:propagation}

Photons traverse an intervening medium some distance before they interact, such that the flux
\begin{align}
\label{eq:fluxdec}
I=I_0 \cdot \exp[-\tau]
\end{align}
follows an exponential decay for the optical depth $\tau$ of the respective interaction process. The optical depth is given by
\begin{align}
  \tau = \int n \sigma dl
\end{align}
along the propagation direction, where $n$ is the number density of the interacting species and $\sigma\left(T,\nu\right)$ is its effective cross-section, which depends on the gas temperature $T$ and the photon frequency $\nu$. As the cross section is evaluated in the frame of the intervening media, the frequency $\nu$ changes due to the gas velocity $v$ as well as cosmological redshifting due to Hubble expansion.

\subsection{Photon Interaction}
\label{sec:interation}

Two major interactions of importance are the destruction and scattering of photons. Destruction refers to absorption followed either by no re-emission, or re-emission with a sufficiently changed frequency that the resulting photon is no longer of interest. In contrast, scattering refers to changes in frequency and/or direction.

For the resonant emission \MgII and \Lya lines, destruction follows interaction with dust, while scatterings can occur following dust interactions (with re-emission in the infrared) or interaction with the ground state of the emission line species.

\subsubsection{Resonant scattering}
\label{sec:scattering}

For convenience, we define the dimensionless velocity $\vec{u} = \vec{v}/v_\mathrm{th}$ and frequency $x=\left(c/v_\mathrm{th}\right)\left(\nu-\nu_0\right)/\nu_0$ for a photon of frequency $\nu$ in a gas parcel with temperature $T$ and velocity $\vec{v}$ for the emission line-center frequency $\nu_0$ and underlying atom mass $m$, yielding a (most probable) thermal velocity $v_\mathrm{th}=\sqrt{2 k_b T/m}$ for the scattering atom mass $m$.

The interacting atom velocity has a large impact on the scattered photon properties. The probability distribution for the two velocity components $u_{\perp_{1,2}}$ perpendicular to the photon direction will follow a thermal Gaussian distribution
\begin{align}
f(u_{\perp_{1,2}}) = \exp{\left(-u_{\perp_{1,2}}^2\right)}
\label{eq:uperp}
\end{align}
Note that the mean is zero as we are in the gas reference frame.

However, the parallel velocity component $u_z$ will follow
\begin{align}
f(u_z) = 
\frac{a}{\pi}
\frac{e^{-u_z^2}}{
    (x_i - u_z)^2 + a^2
} H^{-1}(a, x_i)
\label{eq:uparallel}
\end{align}
given the conditional probability of the interaction in the rest-frame of the atom with a given velocity. To easily describe the interaction, we shift our reference frame via Lorentz transformations in to and out of the atom frame. This changes both the photon frequency and direction. These two transformations are approximated and fused, yielding
\begin{align}
x_{\text{out}} = x_{\text{in}} 
+ \vec{u} \cdot \left(\vec{k}_{\text{out}}-\vec{k}_{\text{in}}\right)
+ g(\mu - 1) 
+ \mathcal{O}(v_{\text{th}}^2 / c^2),
\label{eq:doublelorentztransform}
\end{align}
where velocities are assumed to be small, such that directional changes during Lorentz transformation are negligible. The last two terms represent energy transfer to the atom for conservation of momentum. We ignore this transfer in this work as it is negligible for the scenarios considered here~\citep{Adams71}.

Within the atom reference frame, the direction of the absorbed and quickly ($t\sim A_{\Lya}^{-1}\sim 1\,$ns) re-emitted photon is described by the phase function $p\left(\theta\right)$. We adopt the mix of a 1-to-2 isotropic-dipole phase function near the line center, and a pure dipole for wing scatterings~\citep[$x\gtrsim 0.2$][]{Dijkstra08}.

\subsubsection{Dust}

Photons interacting with dust can be either absorbed or scattered. The relative probability for scattering is set by the albedo
\begin{align}
A = \frac{\sigma_\mathrm{s}}{\sigma_\mathrm{s}+\sigma_\mathrm{a}}
\end{align}
where $\sigma_\mathrm{s}$ and $\sigma_\mathrm{a}$ are the cross-sections for the absorption and scattering by dust. Typical albedo values are between $0.3-0.4$, and we assume a fiducial value of $0.32$ ~\citep{Laursen09}.

For photon scattering, we adopt the effective phase function of~\citet{Henyey41}, given by:
\begin{align}
P_\mathrm{HG}(\mu) = \frac{1}{2}\frac{1 - g^2}{\left(1 + g^2 - 2 g \mu \right)^{3/2}}
\end{align}
where $\mu=\cos\theta$ and $g=\langle\mu\rangle$. We take a fiducial value of $g=0.73$~\citep{Laursen09}.

\subsection{Other emission lines}

For other resonant emission lines, the emission details differ. Generally, we require the species population level, cross-section and phase function to model a given line. For metals, we make use of pre-computed tabulated emissivities from CLOUDY \citep{Ferland17}. This allows us to determine the ioniziation state and metal line emission for each gas cell, under the assumption of collisional plus photoionization equilibrium \citep[as in][see Section~\ref{sec:gible}]{Nelson21}.

Spin-orbit coupling can lead to observable fine-splitting of a transition. While the separation is negligible in the case of \Lya, this splitting results in $\sim 7$\,\angstrom separation of the resonant \MgII doublet~\citep{murphy13}. Each state individually follows the propagation and resonant scattering given its associated optical depth. In such cases of multiple transitions, we assume the overall cross-section equals the sum of the individual values, while re-emission considers the relative branching probabilities. 

In some cases, multiple transitions are relevant but only some are resonant. For example, the \FeII multiplets have a mix of resonant and non-resonant transitions. Upon de-excitation and transition via a non-resonant channel, the photon will have a strongly suppressed chance of interaction due to the low occupation of the lower (excited) transition state, unless close to a resonant transition wavelength.

The physics underlying resonant emission lines -- while not exhaustive in terms of our applications and science -- motivates the overall numerical architecture and design of our new RT code.

\section{thor}
\label{sec:thor}

\subsection{Overview}

\thor is a modern, massively parallel GPU-accelerated MCRT code for radiative transfer. The code is written in modern C++ and uses the SYCL abstraction layer\footnote{\thor supports and is continuously tested against the oneAPI and AdaptiveCpp SYCL implementations with the following target devices: (i) AMD and Intel x86 CPUs; (ii) AMD, Intel, and Nvidia GPUs; and (iii) AMD Instinct accelerator cards.} for comprehensive CPU/GPU vendor support. Multi-node parallelization and inter-node communication is handled via the MPI standard, and we support domain decomposition to enable memory-intensive problems.

SYCL, an open standard maintained by the Khronos Group, is a high-level programming model allowing a single-source code in pure C++ to run on various vendor accelerators and backends. It enables portable, modern, C++17 code while supporting complex data structures, and is an alternative to vendor-specific frameworks such as CUDA~\citep{nvidia-cuda} and ROCm/HIP~\citep{amd-rocm}, as well as libraries such as OpenMP~\citep{OpenMPArchitectureReviewBoard13}, OpenACC~\citep{CAPSEnterprise11}, kokkos~\citep{CarterEdwards14} and RAJA~\citep{Beckingsale19}.\footnote{Each has similar overall goals, with varying paradigms and performance characteristics. We note that several HPC-focused projects including kokkos and RAJA are currently implementing SYCL backends.}

\subsection{Code building blocks}

\thor provides a minimal skeleton architecture and a set of abstract building blocks that allow a diverse set of applications. At the base level, the code is organized around two concepts:
\begin{itemize}
    \item \textbf{Datasets}: Different data formats and geometries are wrapped by a common interface, allowing queries of the underlying source and medium (gas) distributions. We currently implement a uniform grid structure, as well as gridless spherically symmetric and infinite slab geometries.
    \item \textbf{Drivers}: Define the overall program logic, such as the scheduling of the compute kernels, as well as the request for creating and storing Monte Carlo contributions. We currently implement two drivers: resonant line MCRT and raytracing.
\end{itemize}

\label{sec:key_concepts}
The details of the MCRT driver are abstracted in terms of the \textit{interactor}, \textit{generator}, and \textit{output processor}. 
\begin{itemize}
  \item \textbf{Interactors}: Absorption and scattering depend on the simulated emission line (ensemble). The relevant physics is provided by a flexible interactor interface. We currently implement two interactors, namely a single resonant emission line with dust absorption (here, used for \Lya), and a doublet emisison line (here, used for \MgII).
  \item \textbf{Generators}: MCRT photons for a given emission model are spawned by one or more generator instances. Here we implement three generators: a central emission source, a multi-source discrete emission model (e.g. for star/stellar population particles), and a continuous emission model given hydrodynamical (gas) properties.
  \item \textbf{Output Processors}: Finished photons are ingested by one or more output processors, performing required data reduction as needed, and batching write tasks. Implemented processors include: surface brightness maps, spectra, integral-field cubes and raw photon output.
\end{itemize}
The raytracing driver propagates tracer packages through the domain with the  \textit{operator} abstraction enabling customization to computations along the ray.
\begin{itemize}
  \item \textbf{Operators}: Tracer packages primarily hold information such as position, direction, and weight, relevant for their propagation. Custom operators can also be used to passively track additional information or perform additional computations. We currently define four operators that build upon the raytracing driver to compute slices, projections, volume renderings, and absorption spectra (Section~\ref{sec:raytracing}).
\end{itemize}

These abstractions are implemented as C++ concepts and templates. This allows them to be easily combined, and facilities widespread code reusability, while enabling necessary compile-time determination of the device code.\footnote{We use just-in-time (JIT) compilation from the specific SYCL target to accelerate compilation times and device-specific performance.}

\subsection{Datasets and Geometries}
\label{sec:geometries}

\thor can handle different data structures and geometries. The spatial information for hydrodynamical fields is defined by a dataset interface. We currently support three geometries:

\begin{itemize}
\item Spherical Shell: A homogeneous spherical shell of density $\rho$ and temperature $T$, additionally parameterized by its inner and outer shell edge radius $r_\mathrm{i}$ and $r_\mathrm{o}$, and maximum velocity $v_\mathrm{max}$. Velocity can depend on radius, e.g. linearly increasing as $\vec{v}=v_\mathrm{max} \left(\vec{r}/r_o\right)$.
\item Infinite Slab: An infinite slab of a given depth with density $\rho$ and temperature $T$ between $-z$ and $z$, and an optional velocity field $v_\mathrm{z}=v_\mathrm{max} \left(z/r_o\right)$ with maximum velocity $v_\mathrm{max}$.
\item Uniform Grid: A Cartestian grid with uniform spacing and cell size that can hold arbitrary data. We create readers for SPH and Voronoi-based simulations for convenience, as well as a spherical shell interpolation for validation.
\end{itemize}

At a minimum, each dataset needs to provide a single member function \textsc{query} that returns the index of the next cell, the pathlength to that cell, and $N$ hydrodynamical properties $\{q_1...q_\mathrm{N}\}$ requested by the respective driver by compile-time directives. Datasets can implement optional functionality, such as determining the nearest neighbor or handling boundaries, while fallback implementations are chosen at compile-time otherwise.

\subsection{MCRT Driver}
\label{sec:driver_mcrt}

\begin{figure}
    \includegraphics[width=1.0\columnwidth]{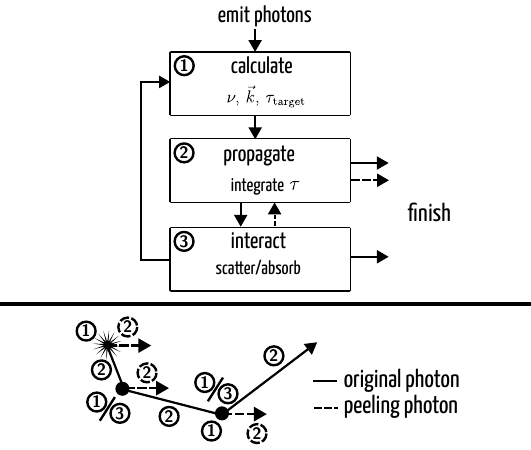}
    \caption{Schematic diagram of the MCRT driver and the process of calculating resonant line emission and scattering in \thor. The source model produces initial photon emission. These photons then propagate through, and interact with, the gas medium.}
    \label{fig:schematic_rt}
\end{figure}

Here, we briefly summarize the implementation of our Monte Carlo radiative transfer (MCRT) approach, following the physics laid out in Section~\ref{sec:rtphysics}. We broadly follow the methodologies of several previous codes \citep[see e.g.][]{Smith15,Michel-Dansac20}. In particular, Figure~\ref{fig:schematic_rt} shows a schematic diagram of the MCRT concept, that we describe below.

\textbf{Emission.} Photons are first spawned by a generator with a given set of weights, frequencies and directions in the rest-frame of the source. Photons are subsequently initialized by shifting into the comoving frame, drawing a propagation optical depth $\tau_\mathrm{target}$, and identifying the starting parent cell index. 

\textbf{Propagation.} Photons subsequently propagate on straight trajectories through the underlying gas distribution until interacting with gas. To attenuate the flux we give each photon a target optical depth $\tau_\mathrm{target}$ randomly sampled from an exponential distribution (Equation~\eqref{eq:fluxdec}). We then integrate the optical depth seen by a given Monte Carlo packet until reaching $\tau_\mathrm{target}$. This signifies an interaction event. Propagation consists of a series of piece-wise linear integrations from cell face to cell face, using \textsc{query} calls to update the integrated optical depth $\tau_\mathrm{current}$ and position.

The optical depth is highly frequency dependent in the case of resonant lines and follows a Voigt profile. We approximate this non-analytical function to efficiently compute the optical depth. In Appendix~\ref{sec:voigt_microbenchmark}, we evaluate different approximations with respect to their performance and accuracy.\footnote{ 
In practice, for \Lya, we use the approximation presented in~\citet{Smith15} that in comparison to the other schemes in~\citet{Michel-Dansac20} provides relative errors below $10^{-4}$ at Voigt parameters $a\lesssim 10^{-2}$. This accuracy is hence always met for \Lya radiation above $T\sim1$\,K. As higher masses and line-center wavelengths imply larger Voigt parameters, such emission lines might require more accurate schemes. We provide a range of compile-time options.}

We assume zeroth order, i.e. constant physical properties within each gas cell, for each propagation step. This can be generalized to higher order in the future~\citep[e.g.][]{Smith25}. In order to capture changes within a cell, e.g. due to the cosmological Hubble flow or a small-scale velocity gradient, we perform sub-cell stepping. Finally, when a photon surpasses $\tau_\mathrm{target}$, we take the interaction site as the linear step path length at which $\tau_\mathrm{target}=\tau_\mathrm{current}$. At this position, we handle the interaction event.

\textbf{Interaction.} Multiple interactions can occur, and across all we compute the total optical depth $\tau$ that defines the distance the photon can travel before the interaction. Each interaction is associated with its own probability $\tau_i$. After a photon is propagated according to $\tau = \Sigma_i \tau_i$ we allow interaction $i$ to occur with a probability proportional to $\tau_i$. Each is handled by an abstract interactor (Section~\ref{sec:key_concepts}). Currently, these treat dust and single- and multi-state resonant emission lines.

For resonant lines in particular, we need to sample the interacting atom velocity. While the perpendicular components are drawn from a Gaussian distribution, the perpendicular direction is drawn from Equation~\ref{eq:uparallel}, which is not analytically integrable. We therefore use rejection sampling~\citep{vonNeumann51} with a piecewise, integrable comparison function \citep[following][]{Zheng02} given by
\begin{align}
g(u_\parallel) \propto 
\begin{cases}
g_1 = \dfrac{1}{a^2 + (x - u_\parallel)^2} & u_\parallel \leq u_0 \\
g_2 = \dfrac{e^{-u_0^2}}{a^2 + (x - u_\parallel)^2} & u_\parallel > u_0
\end{cases}.
\label{eq:uparallel_comparisonfunc}
\end{align}
Different codes implement different critera for $u_0(a,x)$ to minimize the rejection rate and improve overall performance.\footnote{We compare the performance details of several different methods in Appendix~\ref{sec:uparallel_microbenchmark}, where we also discuss future GPU optimization such as pre-sorting of photon data prior to kernel execution. We present a novel method, different from Equation~\ref{eq:uparallel_comparisonfunc} ~\citep[from SK20;][]{Seon20}, that uses ratio-of-uniforms~\citep{Kinderman77} rather than classical rejection sampling outside of the Gaussian core, as well as several classic choices~\citep[S07, L09, S15, D20;][]{Semelin07, Laursen09a, Smith15, Michel-Dansac20}. We find that SK20 (plus S15 within the Gaussian core) gives the highest throughput, while D15 and S20 are comparable.} The choice is selectable at compile-time; we use the acceleration scheme of~\citet{Seon20} in this work.

\subsubsection{Peeling off photons}
\label{sec:peeling}

We implement an optional peeling algorithm to evaluate the emergent radiation field for a given line of sight~\citep{Yusef-Zadeh84,Zheng02,Tasitsiomi06}. In this algorithm, we compute the average luminosity contribution at each scattering that is expected to escape towards an observer. Practically, this requires the evaluation of the phase function along the line-of-sight, multiplied by the exponential flux decay with the optical depth, which must be calculated by integrating a new ray from the point of scattering towards the observer.

In \thor, we log the minimal information (`peel seeds') required for each peeling calculation, in order to evaluate the optical depth during a separate dedicated raytracing computation. As we cannot dynamically allocate memory on the GPU, we pre-allocate a fixed buffer on each device. The buffer is split into a number of peel seed blocks, each able to hold a fixed number of scattering events. In the beginning, each photon is guaranteed an initial block. Once a photon exhausts its block, we spill into yet unused memory by reserving a new block using a global, atomic-locked counter. Photons temporarily cease propagation if no more storage blocks are available. Kernel execution finishes prematurely if no blocks are left.

To further speed up this calculation, we conditionally perform \textit{photon fusing} by combining new contributions with existing peeling photons that have nearly identical properties based on their frequency and position. Heuristic thresholds for fusing are user-specified, based on the required spatial and spectral resolution and the size of resolution elements (Section~\ref{sec:mcst_galaxy}). 

\subsubsection{Core-skipping}
\label{sec:coreskipping}

Core-skipping accelerates the expensive photon propagation due to frequent scatterings near line center. When photons scatter in the core of the line profile within optically thick media, spatial diffusion is negligible. These core scatterings thus do not change observables nor the average photon trajectories and can be skipped. Practically, this is done by a left-truncated Gaussian for a critical dimensionless frequency $x_\mathrm{crit}$~\citep{Dijkstra06}. This idea has been explored in several recent works, and we adopt a scheme with good speed-up and accuracy, choosing $x_\mathrm{crit}=0.2 (a\tau_0)^{1/3}$ for $a\tau_0\geq 1$~\citep[following][]{Smith15}.

\subsubsection{Random numbers}
\label{sec:rng}

MCRT heavily relies on random numbers to sample a range of properties such as emission frequencies, directions and optical depths encountered before interaction. This generally requires pseudo random number generators (PRNGs) with good statistical properties, good performance, and long periodicity.\footnote{Commonly used PRNGs with 32 bit period start showing artifacts, particularly for commonly studied shell models. For example, a static, $\tau_0=10^7$ \Lya shell without core skipping, requires $\sim 10^{15}$ random number draws, compared to a period of $\sim10^{9.5}$ elements.} 

Furthermore, for good performance on GPUs, we require a small RNG state and the generation of independent streams. We provide a range of compile-time RNG choices, and use \textsc{xoshiro128++} by default. Each call to the generator provides 32 or 64 random bits for further use. For each work-item, i.e. photon, we start an independent generator with its sequence and/or state initialized by the MPI rank ID, photon ID, and step ID of the current kernel execution.

We provide a range of floating point random distributions at single/double precision and as unbiased/biased float representations on top of the integer RNG. We can intentionally degrade the precision in parts of the rejection sampling to boost performance while maintaining outcome fidelity. We heavily rely on random float generation $R$ between $[0,1)$. We also implement the other half-open uniform distribution on $(0, 1]$ generated as $1-R$, Gaussian distribution using the Box-Muller algorithm, and random vectors drawn from a sphere using rejection sampling.

\subsection{Raytracing Driver} \label{sec:raytracing}

Raytracing is a general geometrical operation performed on an underlying mesh or grid. It provide integrals or other calculations along finite-length rays that traverse a simulation volume. Our implementation is generic, such that the raytracing kernel accepts a function that is evaluated during each step. This can be used for a variety of purposes, and we discuss three.

\subsubsection{Image Projection}

The projection operation reduces the inherently three-dimensional information of gas properties to a two-dimensional image or map, collapsing (integrating) along a particular direction. The result is informative as a visualization tool, and as an approximate observable that accumulates along a line of sight.

In this case, the kernel function is the line of sight accumulating integral of some quantity $q$, possibly weighted by a second property $w$, as $Q = \int q(s) w(s) ds$ where $s$ is position along a ray. Below we take $q = n_{\rm HI}$ and $w=1$ to obtain column density maps. Normalization by $\hat{Q}=\int w(s) ds$ enables calculation of quantity averages for a given line of sight. For example, with $q = T_{\rm gas}$ and $s = m_{\rm gas}$, we obtain the mass-weighted temperature maps as $Q/\hat{Q}$, i.e. where the image value shows the mean, mass-weighted temperature of all gas along each sightline.

\begin{figure}
    \includegraphics[width=1.0\columnwidth]{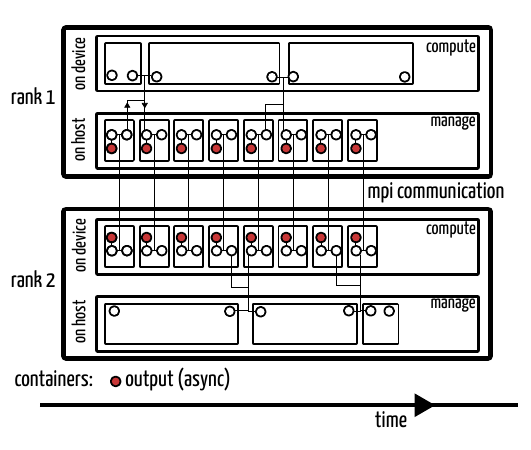}
    \caption{Schematic diagram of the parallelism and communication of \thor. Two distributed-memory MPI ranks are shown. On each, a management thread handles MPI communication, output (red), sorting and pre-processing, and similar tasks. Simultaneously, a compute thread runs driver-specific kernels that execute on the device (CPU, GPU, or APU). The two communicate with thread-safe swaps of double buffers containing photon data.}
    \label{fig:schematic_comm}
\end{figure}

\subsubsection{Synthetic Absorption Spectra}

Closely related to a raytracing-based projection is the creation of a synthetic absorption spectrum. In this case, each parcel of gas intersected along a line of sight deposits optical depth that is distributed in frequency according to a Voigt profile. In the case of \Lya and Equation~\ref{eqn:abs}, this leads to spectral features including Lyman-limit systems (LLSs), Damped Lyman-alpha absorbers (DLAs), and the \Lya forest.

The \Lya absorption forest encodes the imprint of neutral hydrogen along the line of sight path $s$ between a background source with a flux $I_0$ and the observer via
\begin{align} \label{eqn:abs}
\tau_{\Lya}(\nu) = \int n_{\mathrm{HI}}(s) \, \sigma_\Lya\left(\nu(s), T(s)\right) \, ds
\end{align}
leading to a flux $I(\nu)=I_0\exp\left(-\tau_\Lya\right)$ for the given temperature $T(s)$, and neutral hydrogen density $n_\mathrm{HI}(s)$ structure. The frequency $\nu(s) = \nu_i \left[1 - (v(s) + H(z)s)/c \right]$ is set by the peculiar line-of-sight velocity $v(s)$ and the Hubble shift $v_H=H(z)s$.

\subsubsection{Volume Rendering}

Volume rendering is a tool to visualize three-dimensional volumetric data. It renders an image where each pixel obtains its value from an integral along the corresponding ray. Rather than computing a physical quantity, volume rendering maps physical quantities to perceived colors, that can accumulate in unique and informative ways. It uses the integral equation
\begin{align}
\label{eq:volrender}
C = \int_0^L c(s) \, \mu(s) \, \exp\left(-\int_0^s \mu(t) \, dt\right) ds
\end{align}
to obtain the color result $C$ in the image plane, for an inverse ray traversing the volume from the observer at $0$ to $L$. A ray imprints a color $c(s)$ weighted by its absorption coefficient $\mu(s)$, attenuated according to the integrated attenuation to $s$. The color $c(s)$ is commonly specified by a transfer function mapping one or multiple spatial fields $q_i(\vec{x})$ to be visualized to an $\mathrm{RGBA}(q_1, ..., q_N)$, which returns channel float values for red (R), green (G), and blue (B), as well as the alpha (A) channel. The latter sets the absorption coefficient.

\begin{figure*}
    \includegraphics[width=0.95\columnwidth]{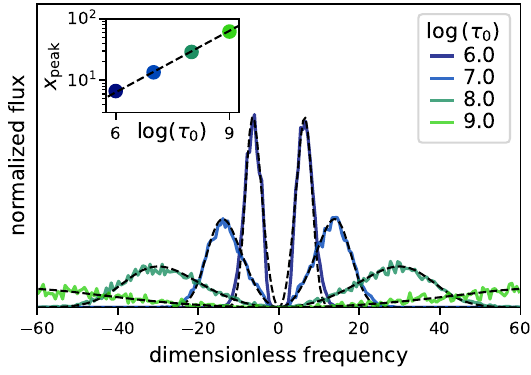}
    \includegraphics[width=0.95\columnwidth]{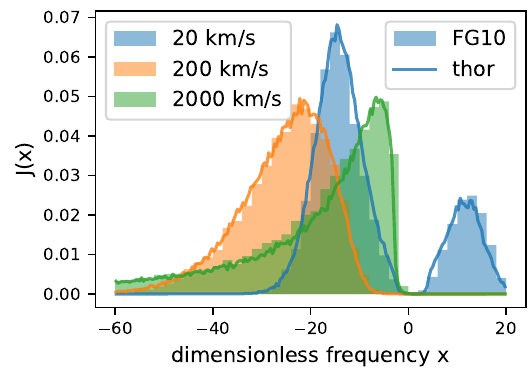}
    \caption{\textbf{Left:} Emergent \Lya spectrum for a uniform density sphere with a line-center optical depth of $\tau=10^6$ and temperature of $T=2 \cdot 10^{4}\,$K. The dashed lines show the analytic solution in the limit $a\tau_0\gg 1$~\citep{Neufeld90,Dijkstra06}. \textbf{Right:} Emergent \Lya spectrum for a homogeneous sphere with a column density of $N_\mathrm{HI}=2\cdot 10^{20}$\,cm$^2$, temperature $T=2\cdot 10^{4}\,$K, and a differential outflow of $v_r(r)=v_\mathrm{max} \left(r/r_\mathrm{max}\right)$. Different colors represent different maximum outflow velocities $v_\mathrm{max}$. The shaded region shows the binned spectrum found by~\citet{Faucher-Giguere10}, while the solid lines show the \thor calculations for which we find excellent agreement.}
    \label{fig:validation}
\end{figure*}

\subsection{Numerical Details}

\subsubsection{Compute kernels and communication}
\label{sec:skeleton}

Each \textsc{thor} process consists of one \textit{compute} thread and one or more \textit{manage} threads. The compute thread schedules and runs driver-specific kernels on a device (CPU or GPU), including the initialization, propagation and interaction of photons. The manage threads handle I/O, communication, load balancing, pre-processing, and related tasks. This scheme is shown in Figure~\ref{fig:schematic_comm}.

Compute and manage threads interact via a photon double buffer, allowing asynchronous operation and keeping the compute device at full load. Each thread has one photon buffer for original and peel photons. Whenever a set of photons has been processed by the managing thread, it is written back to its buffer. Whenever the compute thread finishes its workload, we transfer its result to the CPU and a thread-safe swap of the two photon buffers is executed, and the new buffers are processed. 

The compute thread (and thus each \textsc{thor} process) manages exactly one device. To use multiple devices, across one or more nodes, multiple \textsc{thor} processes can be spawned via MPI. In the case of the MCRT driver, we spawn one manage thread to handle the original photons, one thread to manage peel photons (Section~\ref{sec:peeling}), and optionally one for the generator model.\footnote{Only one compute thread is needed for all photon types as compute kernels can execute asynchronously on the device within SYCL.}

The two main tasks of the manage thread are post-processing of finished photons and communicating photons to other processes as needed. Custom post-processing functions can finalize calculations, before flushing intermediate/final results such as frequency and last-scattering position to disk. This function is batched as an asynchronous thread itself.

The communication of photons is designed to achieve two purposes. First, domain decomposition, whereby we distribute the physical domain across different physical nodes to process large memory-intensive simulation volumes. Second, load balancing, where we balance the workload of multiple processes across one or more nodes to enable efficient multi-GPU, multi-CPU and hybrid CPU-GPU compute workflows.

\subsubsection{Domain distribution}
\label{sec:domaindistribution}

At the beginning of the RT calculation we initialize a domain decomposition object. This divides the domain into non-overlapping subvolumes $\{V_i\}$. Photons are marked as requiring communication (finishing their current kernel execution) if they exit their current subvolume. Each subvolume $V_i$ is padded by a safety margin $d$, overlapping the subvolumes in a way that significantly reduces frequent communication of photons close to the boundary~\citep{Byrohl21}. In practice, we use cuboids for simplicity and leave more complex domain decomposition geometries for future implementation.

When a photon buffer swap occurs, the domain decomposition on each rank determines the next subvolume $V_i$ for every photon in need of communication. In the simplest use case, each subvolume exists exactly once on some MPI rank. In this case, the communication thread constantly makes pairwise photon exchanges with other ranks, which are subsequently processed upon the next buffer swap. However, more complex setups are possible. In particular, we implement a duplication scheme where the same subvolume $V_i$ exists on multiple ranks, letting us perform efficient load balancing for problems without memory pressure.

\subsubsection{Load balancing}
\label{sec:loadbalancing}

We balance workload between MPI ranks for existing photons as well as to-be-spawned photons for a given  generator. For the pool of existing photons, each rank computes a division of photons to each rank holding a given subvolume. based on heuristics from each spawning and receiving rank. Our current implementation guarantees equal photon counts after communication for each rank holding a given domain.

As generators only create photons for the local volume when requested by the driver, their balancing mechanism must work differently. We allow generators to optionally spawn their own communication thread to negotiate and update their generator state according to diagnostic input by the driver, including the current compute buffer size as well as the expected run-time of the current device kernel execution.

\subsubsection{Input and output}

Configuration, including that of physical modules, is set at run-time via \textsc{YAML}-format parameter files. These select options such as output formats and fields, as well as just-in-time compiled device kernels. For output, \thor currently supports \href{https://zarr.dev/}{zarr} and \href{https://www.hdfgroup.org/solutions/hdf5/}{HDF5}. Each MPI rank writes one output file, that can be later merged. File writes are batched as asynchronous threads during other host and device processing tasks (Figure~\ref{fig:schematic_comm}).

\thor supports various outputs including raw photons, surface brightness maps, and spectra. Output processors can signal for the completion of a RT calculation, e.g. for run-time determination of MCRT convergence in an observable output statistic. This increases ease of use, as \thor can directly output final science results that are both automatically robust and highly configurable.

\section{Validation}
\label{sec:validation}

We begin by validating the correctness of \thor. We focus on end-to-end tests of the \Lya emission line for this purpose.

Figure~\ref{fig:validation} shows the commonly studied \Lya `Neufeld' regression test (left panel) for a homogeneous sphere with different line-center \Lya optical depths. Photons are injected at line-center at the origin. We consider $\log(\tau_0) \in \{6, 7, 8, 9\}$, and in all cases the medium has a temperature of $T=10^{4}\,$K. We compare the numerical results of \thor with the analytic solution, that is valid only in the limit of extremely high optical depth~\citep{Neufeld90,Dijkstra06}. For this test we use our meshless spherical shell geometry (Section~\ref{sec:geometries}), and do not explicitly realize a discretized gas distribution. We do not use the core-skipping acceleration scheme.

The emergent \Lya spectrum evolves from a narrow double peak with small separation, to a broad double peak with wide separation (as shown in the inset, in terms of the dimensionless frequency $x$). We find good agreement between \thor (solid colored lines) and the analytic solution (dashed black lines), particularly at high $\tau_0$ where the assumption for the analytic solution $a\tau_0\gg 1$ is fulfilled.

Figure~\ref{fig:validation} also shows the results for a second spherical setup with $N_\mathrm{HI}=2\cdot 10^{20}$\,cm$^{-2}$ and $T=2\cdot 10^{4}\,$K, now including a differential outflow velocity $v_r(r)=v_\mathrm{max} \left(r/r_\mathrm{max}\right)$ (right panel).  The emergent \Lya spectrum depends strongly on the velocity scaling parameter, evolving from a red-dominant double peak (20 km/s) to strongly asymmetric single peak profiles (200 km/s and 2000 km/s). 

While no analytical solution exists, we validate against previous numerical calculations~\citep[solid histograms;][]{Faucher-Giguere10}, finding good agreement with the \thor calculations (solid lines). This end-to-end test of the analytic sphere geometry with non-trivial velocity gradient validates many details of the MCRT treatment.

\begin{figure}
    \includegraphics[width=0.95\columnwidth]{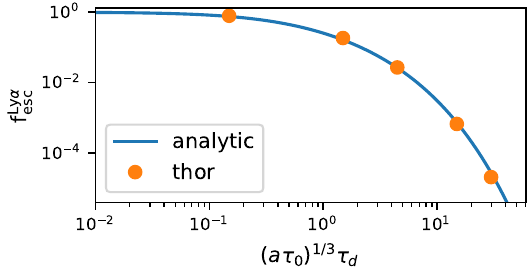}
    \caption{Validation of \Lya escape fraction for photons injected in the line center within an infinite slab. Parameterized by $\left(a\tau_0\right)^{1/3}\tau_a$ an analytic solutions exists for $a\tau_0\gg 1$ and $a\tau_0\gg \tau_a^3$~\citep{Laursen10}. We find good agreement between the \thor calculations and the analytic expectation across the full parameter range.}
    \label{fig:fesc_vs_tau_dust}
\end{figure}

\begin{figure}
    \includegraphics[width=0.95\columnwidth]{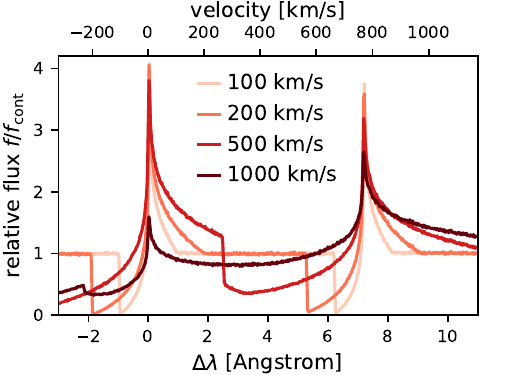}
    \caption{Validation of the doublet implementation for \MgII. We initialize a sphere of constant density with a differential velocity gradient as for \Lya, with $\mathrm{N}_\mathrm{MgII}=10^{14.5}$\,cm$^{-2}$ and $T\sim 10^4$\,K, and consider four different outflow velocities $v_\mathrm{outflow}$. An initially flat continuum is sampled by $10^7$ photon packets, and the resulting emergent spectrum is shown. As we increase the outflow velocity, the peak ratio shifts towards the lower frequency peak, particularly once the outflow exceeds the line separation of $\sim 770$\,km/s. The result can be quantitatively compared against the numerical results of~\citet{Chang24}.}
    \label{fig:mgii_shell}
\end{figure}

Next, we test our dust implementation. Figure~\ref{fig:fesc_vs_tau_dust} shows the results of a test problem where we inject \Lya photons at line center, within a spatial geometry that is a finite thickness, infinite slab of gas. An analytic solution exists for the escape fraction $f_\mathrm{esc}=L_\mathrm{esc}/L_\mathrm{int}$ where $L_\mathrm{int}$ is the intrinsic (emitted) luminosity, and $L_\mathrm{esc}$ is the (emergent) luminosity escaping towards the observer. We compute $f_\mathrm{esc}$ for several different values of $b=(a\tau_0)^{1/3}\tau_d$ where $\tau_0$ and $\tau_d$ are the \Lya line center and dust optical depth to the slab center, respectively. We adopt a fixed $T=20\,$K and $\tau_0=10^6$ and vary $\tau_d=0.01, 0.1, 0.3, 1.0, 3.0$ (orange markers). In comparison, the analytic solution for $f_\mathrm{esc}(b)$~\citep[blue line;][]{Laursen09} is in excellent agreement with our numerical calculation.

In order to test and validate our ability to run more complex interactors, we implement the \MgII doublet. Figure \ref{fig:mgii_shell} shows the results of a simple test problem. We initialize a sphere of constant density and differential velocity gradient, similar to the previous test for \Lya. We choose $N_\mathrm{MgII}=10^{14.5}$\,cm$^{-2}$ and $T\sim 10^4$\,K for different outflow velocities $v_\mathrm{outflow}\in\left\{100,200,500,1000\right\}$\,km/s.

For each of the doublet states, we see a characteristic P Cygni profile, i.e. a blueward absorption feature in the continuum next to the emission peak, typical for resonant lines within outflowing geometries~\citep{prochaska11}. The width of the absorption feature is directly tied to the maximum velocity difference within the sphere (here: $100-1000$\,km/s). The sharp onset of absorption can be seen for a line shift corresponding to the respective maximal velocity difference.

For $1000$\,km/s, the velocity exceeds the peak separation. In this case, the otherwise invariant emission peak ratio between H and K state changes because photons from the K state eventually re-scatter into the resonance of the H state in the outskirts of the sphere. We compare our results to previous numerical calculations of the same setup and find excellent agreement ~\citep[][their Figure 21]{Chang24}.

Additionally, we perform a range of unit tests (not shown) to validate the behavior of physical components, such as the Voigt profile $V(x,a)$ and core-wing transition $x_\mathrm{cw}$ (see Appendices~\ref{sec:voigt_microbenchmark} and \ref{sec:xcw}). We specifically validate probabilistic events e.g. when sampling the $u_\parallel$ velocity distributions (see Appendix~\ref{sec:uparallel_microbenchmark}), which is closely tied to the random number quality.\footnote{Our adapted RNG generation performs significantly better than other commonly used RNG generators in \Lya MCRT, some of which induce RT artifacts. We confirm the random number generation in distributed and massively parallel RNG states, and verify the underlying random number streams with the well-established \href{https://pracrand.sourceforge.net/}{PractRand} suite.}

\section{Scientific Use Cases -- Showcase}
\label{sec:applications}

Next, we present several scientific use cases of \thor, highlighting its wide applicability. We broadly order these applications by spatial scale. First, we explore parameter inference from \Lya spectra via simplified shell geometries (Section~\ref{sec:parametersweep}), \Lya signatures in high-resolution, high-redshift galaxies (Section~\ref{sec:mcst_galaxy}), \MgII and \Lya signatures of the circumgalactic medium (Section~\ref{sec:gible}), and cosmological boxes (Section~\ref{sec:cosmo}).\footnote{Throughout we use $\log$ to indicate the logarithm at base $10$ and $\ln$ for the natural basis. For cosmological simulations and their analysis, a Planck 2015 compatible cosmology with $\Omega_{\Lambda,0}=0.6911$, $\Omega_{m,0}=0.3089$, $\Omega_{b,0}=0.0486$, $\sigma_8=0.8159$, $n_s=0.9667$ and $h=0.6774$ is assumed~\citep{PlanckCollaboration16}.}

\subsection{Lyman-alpha Shell Model Parameter Exploration}
\label{sec:parametersweep}

The interpretation of \Lya spectra often employs the use of large sets of radiative transfer simulations in simplified geometries for physical parameter inference~\citep[e.g.,][]{Gronke15,Gurung-Lopez22,Garel24}.
Here, we demonstrate the potential of \thor for this application. To do so we combine a large grid of idealized RT runs with Markov chain Monte Carlo (MCMC) sampling to obtain the posterior distribution that best fits a given spectrum.

Our setup is as follows. First, our meshless geometry (Section~\ref{sec:geometries}) supports different power-law configurations of neutral hydrogen density $\rho(r)\propto r^{\alpha_\rho}$ and velocity with $v(r)\propto r^{\alpha_v}$. We normalize velocity so that the distance-averaged value equals a given $v_\mathrm{avg}$ within the shell, and normalize density such that the column density between inner and outer shell equals a given total $N_\mathrm{HI}$. The shell width is chosen from a velocity ratio $\Delta_v=\max\left(r_i/r_o, r_o/r_i\right)$ between the inner ($r_i$) and outer ($r_o$) shell radius. Along with the temperature $T$, this gives six parameters.

We then perform `post-processing RT' to continuously vary the intrinsic spectrum Gaussian width $\sigma_\mathrm{HI}$, the equivalent width $\mathrm{EW}$, and the dust optical depth $\tau_d$. This is achieved by rescaling the scattered photons, using their saved initial and final frequencies and $N_\mathrm{HI}$ traversed \citep[following the methodology of][]{Gronke15}. Before rescaling, we assume an intrinsic equivalent width of $3$\angstrom and width of $800$\,km/s, and simulate the continuum within $\pm 2500\,$km/s around the line center. For each RT run, we spawn $2\cdot 10^5$ photons. To capture velocity and density gradients, we impose a global step limiter such that the maximum step size $l_\mathrm{max}=f\cdot \min\left(l_v, l_\rho\right)$ where $l_v=v_\mathrm{th} / \max(dv/dr)]$ and $l_\rho=\max[\rho / (d\rho/dr)]$ with $f=0.01$, i.e. a photon can never traverse a length within which the velocity change in terms of the thermal velocity or the local relative density change exceeds $1\%$~\citep[also see][]{Smith22a}. Runs are performed without any core-skipping scheme.

\begin{table}[h!]
\centering
\begin{tabular}{ll}
\hline
\textbf{Param} & \textbf{Values} \\
\hline
$v_{\mathrm{avg}}$ & 80, 90, 100, 110, 120, 130, 140, 150, 160 \\
$\log\left(N_{\mathrm{HI}}/\mathrm{cm}^{-2}\right)$ & 17.6, 17.8, 18.0, 18.2, 18.4, 18.6, 18.8 \\
$\log\left(T/\mathrm{K}\right)$ & 3.0, 3.5, 4.0, 4.5, 5.0 \\
$\alpha_v$ & 0.0, 0.25, 0.5, 0.75, 1.0, 1.25, 1.5 \\
$\alpha_\rho$ & 0.0, $-0.25$, $-0.5$, $-0.75$, $-1.0$ \\
$v_{\mathrm{ratio}}$ & 1.0, 1.5, 2.0, 3.0, 4.0, 5.0, 10.0, 15.0, 20.0 \\
$\sigma$ & continuous \\
EW & continuous \\
$\tau_d$ & continuous \\
\hline
\end{tabular}
\caption{Parameter values used in the \Lya shell model parameter exploration study. The first six are explicitly realized by RT runs, while the last three are achieved in post-processing. The total parameter space has a dimensionality of nine.}
\label{tab:parametersweep}
\end{table} 

Table~\ref{tab:parametersweep} shows the parameter space of the RT grid that we carry out. Using these results, we then select a single mock (i.e. observed) spectrum, with the parameters $\log\left({N_{\mathrm{HI}}/\rm{cm}^{-2}}\right)=18.4$, $v_{\mathrm{avg}}=120$\,km/s, $\log\left(T/\mathrm{K}\right) = 4.0$, with an intrinsic Gaussian width of $\sigma_{\mathrm{HI}}=300$\,km/s, equivalent width $\mathrm{EW}=100$\,\angstrom, and dust optical depth $\tau_d=0.2$. The velocity and density gradients are chosen as $\alpha_v=1.0$ and $\alpha=-0.5$ with a velocity ratio $\Delta_v=10.0$. We add Gaussian noise with $\sigma=0.2$ times the mean flux to each spectral bin. Spectra are binned with $\Delta\lambda_\mathrm{restframe}=0.2$\,\angstrom ($R\sim 6000$). We evaluate the quality of fit by the $\chi^2$ with respect to the normalized mock spectrum. We use \textsc{emcee}~\citep{Foreman-Mackey13} to sample the posterior distribution, running $300$ chains of $3000$ steps each with a Gaussian log-likelihood. 

\begin{figure*}
    \includegraphics[width=0.95\textwidth]{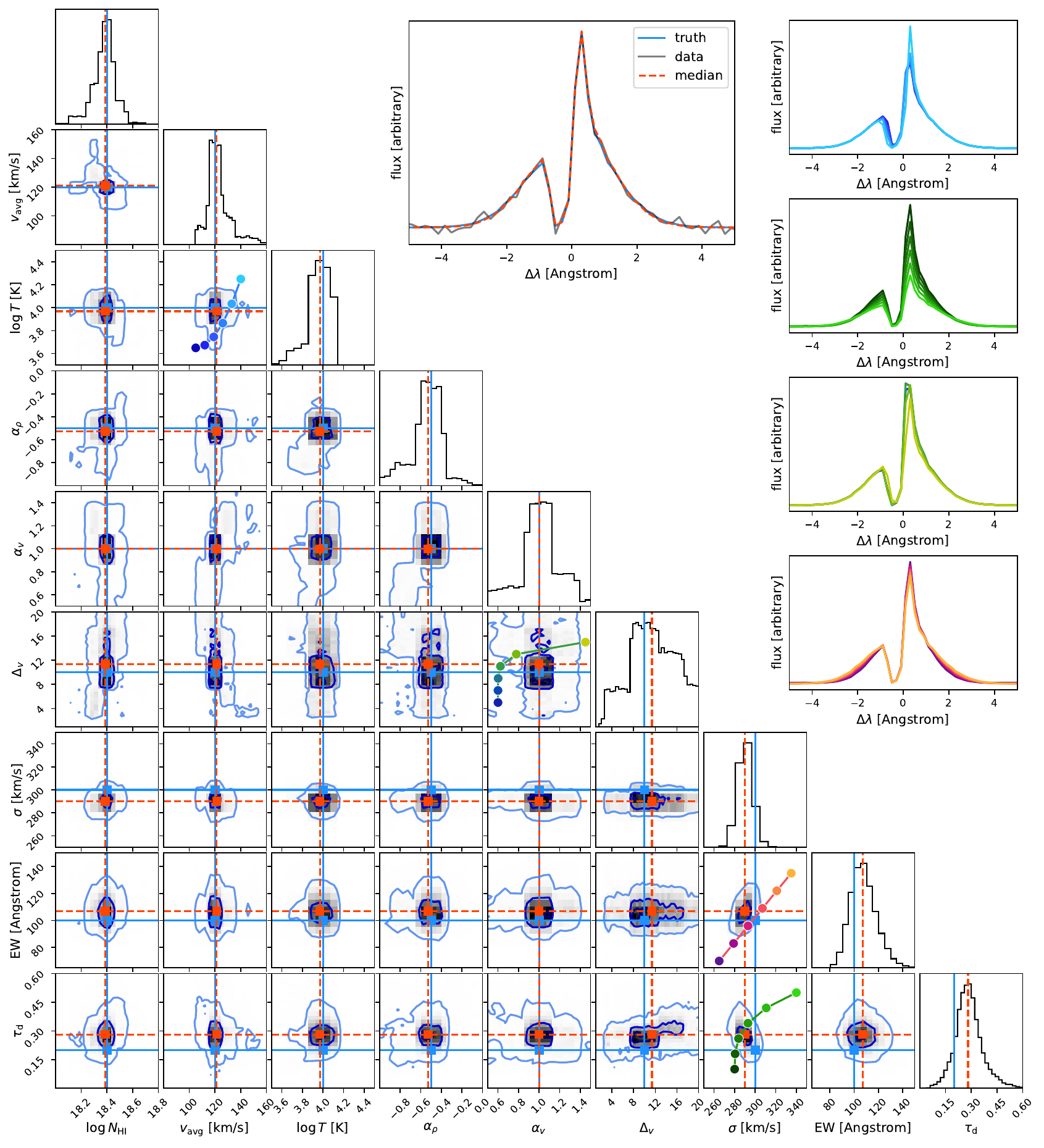}
    \caption{Corner plot of the posterior distribution for $\log_\mathrm{HI},\, v_\mathrm{max},\,\log T,\,\sigma_\mathrm{int},\,\mathrm{EW}_\mathrm{int},\,\tau_d$ from \Lya mock spectrum using $\sim 50,000$ \thor RT simulations, and \textsc{emcee} for sampling. Blue contours show the $0.5\sigma$, $1\sigma$, and $2\sigma$ levels. In the upper right corner, we show the true (i.e. `observed') spectrum, and the posteriori median realization. The smaller left panels show how the emergent spectra depend on parameter variation, across the four particular colored trajectories added to the corner plot panels themselves.}
    \label{fig:cornerplot}
\end{figure*}

Figure~\ref{fig:cornerplot} shows the resulting corner plot of the posterior distribution, across the nine model parameters. The large upper panel shows the true spectrum (blue), its noisy mock (gray), and the median of the posterior (red). The four smaller panels on the right show how the resulting \Lya spectrum responds to parameter variation, along trajectories through the parameter space as shown in the corner plot itself. Successful recovery of a good fit demonstrates how \thor can be used to generate and use large suites of idealized geometry RT calculations.

However, while simplified geometries for spectral inference are tractable, their physical meaningfulness is limited \citep{nianias25}. In addition to degeneracies within the parameter space, it is unclear how the inferred parameters correspond to physical properties of the ISM or CGM. Finally, the highly anisotropic morphology of true galaxies casts doubt on a one-dimensional treatment. Even a single galaxy, at one moment in time, can give rise to a large diversity of observed spectra depending on the line-of-sight~\citep{Blaizot23}.

\subsection{High-redshift Lyman-Alpha Emitting Galaxy}
\label{sec:mcst_galaxy}

\begin{figure*}
  \includegraphics[width=0.98\textwidth]{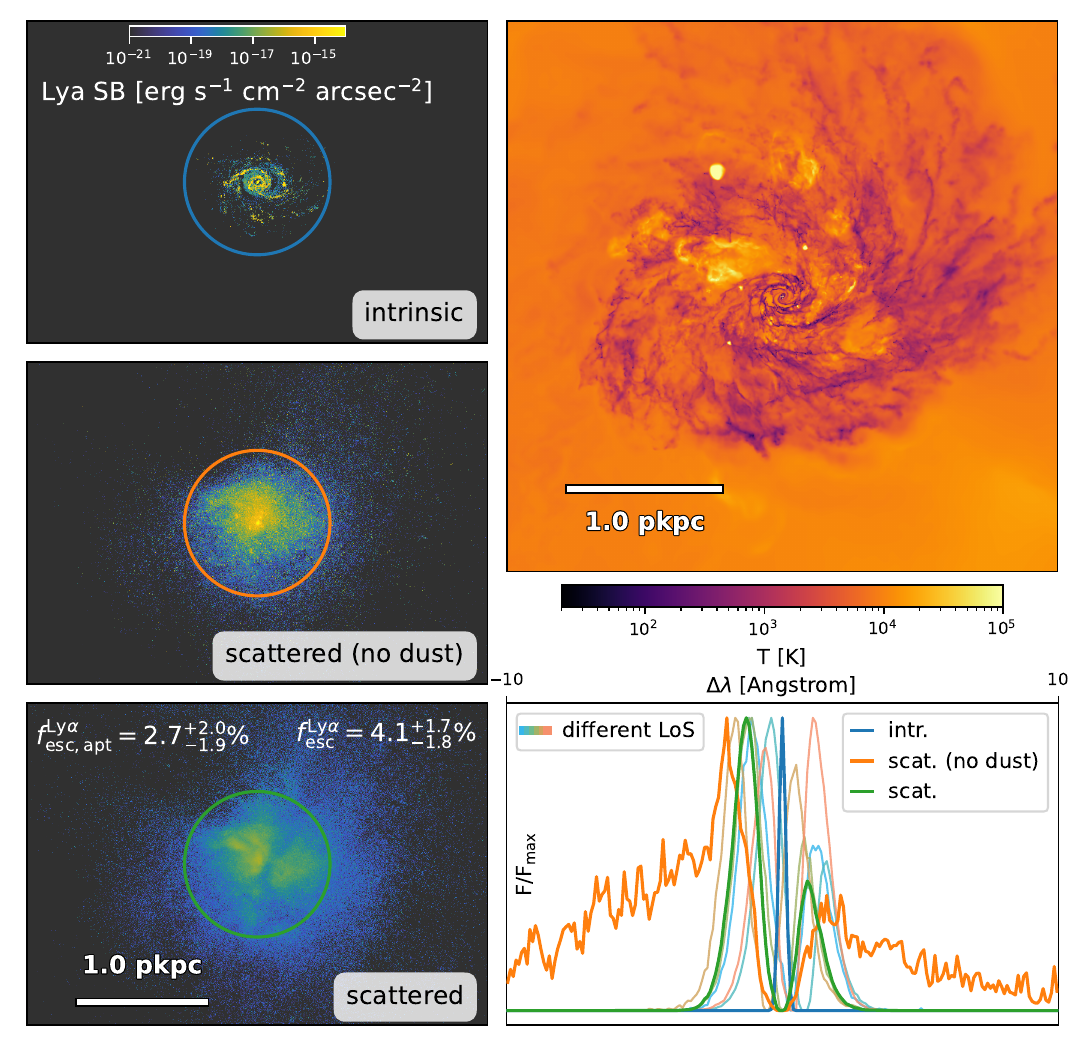}
  \caption{Application of \thor raytracing and \Lya MCRT for a highly resolved, high-redshift galaxy formation simulation at $z=6$. \textbf{Left column:} \Lya surface brightness maps. The top panel shows the intrinsic emission from recombinations due to local stellar ionizing photons. The middle panel shows the surface brightness map after resonant scattering if no photon destruction by dust occurs. The third panel shows the SB map when including dust, resulting in significant central dimming. \textbf{Right column:} the top panel shows a zoom into the central galaxy showing a projection of mean gas temperature, computed with the \thor raytracing operator. The bottom panel shows the \Lya spectra extracted from a central $r=0.5$\,arcsec aperture, contrasting the three cases (intrinsic, dust free, and with dust).}
  \label{fig:mcrt_galaxy}
\end{figure*}

Hence, we next consider \Lya radiative transfer on top of a realistic, three-dimensional galaxy within a high-resolution cosmological galaxy formation simulation at $z=6$. As we use simulations run with the \textsc{AREPO} code, individual finite volumes of gas are represented by Voronoi polyhedra (cells). For our applications here, we interpolate this data onto a uniform grid via a standard SPH mapping \citep[following][]{nelson16}. Each Voronoi cell is deposited onto the grid with the usual 3D cubic spline SPH kernel~\citep{Monaghan92} with a kernel size of $l_i=2.5\cdot \sqrt[3]{3V_i/4\pi}$, where $V_i$ is the Voronoi cell volume. Densities, temperatures, and velocities are weighted such that mass, energy, and momentum are conserved during interpolation. We map to a $3072^3$ uniform grid that spans the host halo virial radius, yielding a $\sim 1$\,pc resolution. While not sufficient to capture the full small-scale detail of this simulation, it is adequate for the present demonstration.

This simulation is part of an upcoming high-resolution galaxy simulation suite (\textcolor{blue}{Nelson et al. in prep}) based on a resolved ISM and explicit stellar feedback physics model (\textcolor{blue}{Smith in prep}). This includes processes such as photoionization of HII regions, photoelectric heating, time-resolved individual supernovae, stellar winds, individual IMF sampling for massive stars~\citep[extending][]{Smith21,Smith21a}. Cooling uses a non-equilibrium primordial chemical network coupled to tabulated metal cooling tables including self-shielding from an incident time-varying uniform UV background \citep{faucher20}. The simulation has a gas and stellar mass resolution $m_\mathrm{baryon} \simeq 24$\msun, with corresponding softening lengths around $\sim 1$\,pc.

Figure~\ref{fig:mcrt_galaxy} shows a detailed, close-up view of the galaxy itself, in projected gas temperature (upper right panel). In this nearly face-on view we see substantial small-scale, spiral-like structure, that is also present in the cool neutral gas. Individual supernovae explosions can be seen heating gas in their vicinity to $>10^5\,$K.

For \Lya post-processing, we assume photons are produced proportional to the ionizing photon rate associated with each star, under the assumption of case-B combination at fixed temperature ($T=10^{4}$\,K) as $L_{\mathrm{Ly}\alpha}=0.68 E_{\mathrm{Ly}\alpha} \dot{N}_\mathrm{ion}$. Photons are injected following a Gaussian distribution with $\sigma=30\,$km/s in the rest-frame of the gas. 
Per \Lya source, we spawn $10^{7}$ photons per $10^{42}$\,erg/s luminosity with a minimum (maximum) photon count of $10$ ($10^{3}$).
We use our adaptive core-skipping scheme, and fuse peeling photons with separations below $0.01$\,\angstrom and $0.3$\,pc. Dust is modeled with a number density $n_\mathrm{dust}=\left(n_\mathrm{HI}+f_\mathrm{ion}n_\mathrm{HII}\right)Z/Z_0$ with the SMC dust cross-section~\citep[following][]{Laursen09}. We adopt $f_\mathrm{ion}=0.01$ as is common~\citep[see][for a discussion]{Smith22}.

Figure~\ref{fig:mcrt_galaxy} shows the resulting \Lya surface brightness maps for the intrinsic emission (top left), scattered photons when ignoring dust (middle left), and scattered photons when considering dust (bottom left). We also show the respective emergent spectra within a $r=0.5\,$pkpc aperture (bottom right). The intrinsic emission reveals that ionizing radiation is concentrated in the innermost $< 1$\,pkpc region of the galaxy, even though the disk is much more extended. Subsequently, \Lya radiative transfer redistributes photons into the more extended neutral hydrogen. The effect of dust is to obscure the central peak in the surface brightness map, leading to a biconical/X-shaped morphology with a centrally dim channel.

The intrinsic emission spectrum is confined to the line center (blue line in the inset panel). After RT, however, a strong double-peaked, blue dominant feature is present. Without dust, long tails extend more than $10$\,\angstrom (rest-frame) from the line-center (orange line). However, the presence of dust limits the spectral width to a few Angstrom around the line center~\citep[green line, confirming][]{Laursen09}. This reflects the higher HI column density, and hence, dust column density, experienced by photons far from the line center, that leads to their destruction. Moreover, the inset spectra panel shows six different lines of sight. We find an impressive diversity in the resulting line shapes \citep[in line with][]{Blaizot23}, with varying peak separation and peak asymmetry. The peak asymmetry, the blue flux over the total flux $r_\mathrm{blue}$ within the aperture, is commonly $\sim 0.7$. However, one line of sight has a mild red flux excess ($r_\mathrm{blue}=0.46$).

We furthermore compute the Lyman-alpha escape fraction $f_\mathrm{esc}^{Ly\alpha}$ for each line of sight, as well as the escape fraction restricted to the $r=0.5$\arcsec aperture $f_\mathrm{esc,apt}^{Ly\alpha}$. The latter is more comparable to observational measurements, e.g. as now commonly being made with JWST in order to better constrain the sources of cosmic reionization \citep{alek24,lin24,chen24,tang24}. To do so we divide the emergent flux by the intrinsic flux (i.e. without RT and dust). We find $f_\mathrm{esc}^{Ly\alpha}={4.1}^{+1.7}_{-1.8}\,\%$ and $f_\mathrm{esc,apt}^{Ly\alpha}={2.7}^{+2.0}_{-1.9}\,\%$. The ranges show the diversity of the six sampled lines of sight, reflecting significant variability on ISM scales. The difference between the two escape fraction shows the significance of \Lya scattering out of the aperture. Typically observationally inferred escape fractions are associated with the \Lya escape itself, while our results demonstrate the importance of scatterings, potentially leading to a systematic over-prediction by a factor of 2 for the \Lya escape.

\subsection{Illuminating the Circumgalactic Medium}
\label{sec:gible}

\begin{figure*}
  \includegraphics[width=0.98\textwidth]{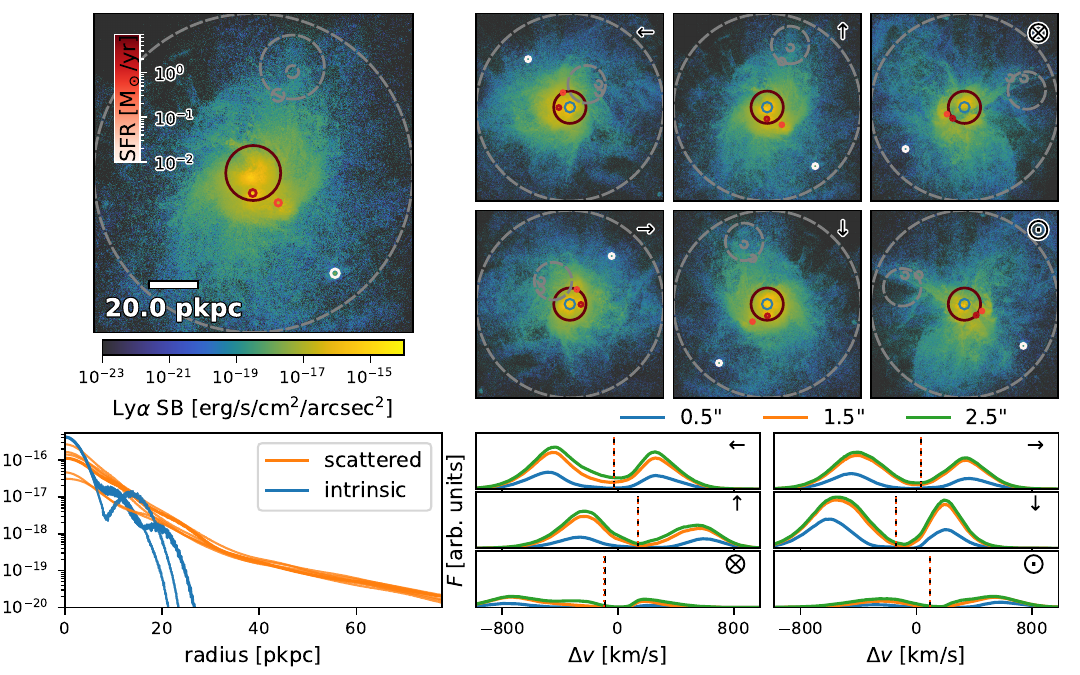}
  \caption{Impact of resonant scattering on \Lya emission in the circumgalactic medium of the highly-resolved cosmological hydrodynamical galaxy formation simulation GIBLE at $z=2$ (S167RF4096). \Lya is emitted proportional to the star-formation rate of the galaxy, and its satellites, as point sources. \textbf{Top left}: surface brightness map of emergent \Lya emission, with a FWHM PSF of $0.2$\,arcsec. \textbf{Bottom left}: Radial surface brightness profiles, including a PSF of $0.7$\,arcsec FWHM.
  \textbf{Top right:} \Lya surface brightness maps along six different orthogonal viewing directions, all random with respect to the central galaxy. We show the $r=0.5$\,arcsec aperture in blue. \textbf{Bottom right:} The corresponding \Lya spectra for the same six directions. The spectra are evaluated in the cosmological rest frame, and the flux normalization is the same in all panels. The vertical black (red) line shows the line-of-sight velocity shift of the halo (galaxy). The total aperture flux strongly varies with the line of sight.}
  \label{fig:gible_lya}
\end{figure*}

We now consider extended emission signatures of the resonant \Lya and \MgII lines at $z=2$ and $z=1$, respectively, where both lines trace the cool gas phase at $\sim 10^{4}$\,K. In particular, extended \Lya emission is an ubiquitous feature of Lyman-alpha emitters~\citep{Wisotzki18,LujanNiemeyer22}, high-redshift galaxies~\citep{Leclercq17,LujanNiemeyer22a} and quasar host halos \citep{borisova16,cai19}, readily observable with ground-based instruments at $z\gtrsim 2$. With \thor, we can forward-model these observables from different models.

Simulations and radiative transfer studies show that \Lya signatures trace the complex gas structure of the CGM~\citep{Zheng11a,Yajima13,Lake15,Byrohl21, Mitchell21}. Here, we take a cosmological magnetohydrodynamical zoom simulation of the progenitor of a Milky Way-like galaxy at $z=2$ from the GIBLE project \citep{Ramesh24,Ramesh24a}. The simulation achieves a $225$\msun resolution in the CGM, allowing us to resolve cold gas clouds down to this mass scale -- the median spatial resolution is $\sim 40-400$ physical parsecs in the CGM, depending on distance.

Figure~\ref{fig:gible_lya} shows the outcome of resonant \Lya RT for a realistic $z=2$ cosmological CGM from the GIBLE suite.\footnote{We re-grid the Voronoi gas distribution onto a uniform Cartesian mesh using SPH splatting with a resolution of $3072^3$ spanning the virial radius $\pm 77.5$\,pkpc in side-length, giving a spatial resolution of $\sim 50$\,physical parsec. The $z=2$ target halo has a mass of $4.5\cdot 10^{11}$\msun.} We adopt a simple emission model that injects photons in the center of each galaxy, with intrinsic luminosity linearly proportional to the galactic star formation rate as
\begin{align}
\label{eq:sfrlya}
L_\mathrm{Lya} = f_\mathrm{eff}\frac{\mathrm{SFR}}{\mathrm{M}_\odot/\mathrm{yr}} \times 10^{42} \,\mathrm{erg/s}
\end{align}
We therefore assume that a fraction of the ionizing photons emitted from stars subsequently lead to recombinations giving rise to \Lya photons. The proportionality factor $f_\mathrm{eff}$ depends on modeling assumption, but is of order unity~\citep{Dijkstra17}. For each \Lya source, we spawn $3\cdot 10^{5}$ photons per $10^{42}$\,erg/s luminosity with a minimum (maximum) photon count of $10^{4}$ ($10^{7}$).

Figure~\ref{fig:gible_lya} shows the resulting CGM-scale emission after \thor \Lya scattering MCRT. In the upper left, we show the surface brightness map for the scattered photons. Intrinsic emission from the central galaxy and satellite galaxies (marked with SFR-colored circles)\footnote{There are three star-forming galaxies within the target halo, with the two innermost galaxies having similar stellar masses of $3-6\times 10^9$\msun and star-formation rates of $7.1$ and $2.3$\msun/yr respectively. While we consider gas cell contributions from neighboring halos, we do not include emission from subhaloes associated with other halos.} scatters and illuminates the hydrogen distribution in the CGM, producing detectable surface brightness levels of $\geq 10^{-18}$\sbunits.

In the upper right, we show the emergent surface brightness (SB) maps for the six different random lines of sight. The detailed morphology of the \Lya emission depends on viewing direction, although the overall SB levels are similar. We quantify this with the corresponding radial surface brightness profiles, within a $r=0.5\,$arcsec aperture (lower left panel). There is variation in both intrinsic and scattered/observable profiles, where some lines of sight have significantly higher SB values, particularly within $\lesssim 30$\,pkpc. The corresponding spectra are shown in the lower right panel, for 0.5\arcsec (blue), 1.5\arcsec (orange), and 2.5\arcsec (green) apertures. A clear double peak with a peak separation of $700\,$km/s emerges in most sightlines, typically with a mild excess in the red peak. The flux normalization is the same in all six cases, i.e. some sightlines show $\sim 7$ times as much flux as others.

\begin{figure*}
  \centering
  \includegraphics[width=0.98\textwidth]{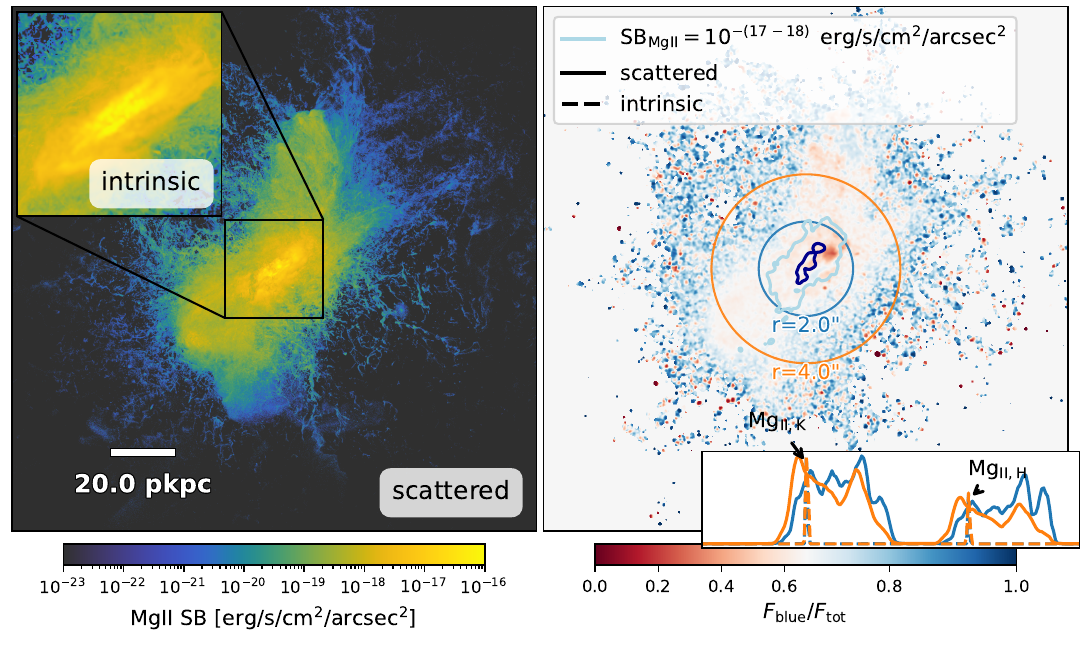}
  \caption{Extended \MgII emission on the scales of the circumgalactic medium of a star-forming galaxy at $z=1$ from the GIBLE simulations. \textbf{Left}: surface brightness map after accounting for scattering. The inset shows the intrinsic emission in the inner region. \textbf{Right:} Spectral properties, showing the doublet peak ratio F$_\mathrm{blue}$/F$_\mathrm{tot}$ map in the image plane. We evaluate the spectrum in the comoving frame at the position of the galaxy and consider those contributions "blue" for which $\lambda < \lambda_c=\left(\lambda_{\mathrm{MgII,K}} + \lambda_{\mathrm{MgII,H}}\right)/2$. The inset shows the annulus spectrum within given aperture radius. Overall, there is a mild excess of MgII,H. Next to the central galaxy, we find some moderately offset red (Mg$_\mathrm{II,H}$) excess, while outskirts appear to be spatially dominated by a blue (Mg$_\mathrm{II,K}$) excess.}
  \label{fig:cgm_mgii}
\end{figure*}

Next, we turn our attention to emission from diffuse \MgII haloes, as recently detected in observations~\citep{Zabl21,Leclercq22,dutta23}. We use the $z=1$ snapshot of the same Milky Way progenitor from the GIBLE simulation suite. We adopt a substantially different emission model, where each gas cell is a source. We compute the \MgII ionization state and emissivity in post-processing using CLOUDY~\citep{Ferland17}, following~\citet{Nelson21}. \MgII emission is injected as a spectral delta peak for each gas element. The relative weight for each doublet state is proportional to its oscillator strength. For efficiency, we limit emission of photons of gas cells with a total luminosity of $> 10^{33}$\,erg/s. For each emitting gas cell, we spawn $10^{6}$ photons per $10^{42}$\,erg/s luminosity with a minimum (maximum) photon count of $10$ ($10^{3}$).

The left panel of Figure~\ref{fig:cgm_mgii} shows the scattered \MgII surface brightness map. For comparison, the inset shows the intrinsic emission from the central region, that is concentrated within the central galaxy. However, photons then diffuse substantially in space, tracing out cool CGM gas. The scattered/observable surface brightness map has qualitatively different characteristics than the \Lya maps. In particular, large \Lya optical depths along most lines of sight produce significant dimming, while the more moderate \MgII column densities sharply trace cold hydrogen clouds, filaments, and structures in the CGM.

In the right panel of Figure~\ref{fig:cgm_mgii}, we show observable spectral diagnostics: the spatially resolved map of peak ratio, and the aperture-integrated emergent vs. intrinsic spectra for the \MgII doublet.\footnote{Since we only consider gas emission (and no stellar continuum), absorption features blue of a given emission are not present (compare with Figure~\ref{fig:mgii_shell}).} We find non-trivial spatial variations of the peak ratio, with a red-dominant central region, while the outskirts are blue-dominant. and an integrated asymmetry of the peak ratio towards the \MgII H state.

The lower inset shows the spectra themselves, for a $r=2$\arcsec\xspace aperture (blue), $r=2-4$\arcsec\xspace annulus (orange), and $r=4-8$\arcsec\xspace annulus (green). In all three cases, the dashed lines show intrinsic emission, that is broadened due to gas velocities in the galaxy. In contrast, the scattered emission is qualitatively transformed. There is (1) an overall velocity shift by $\sim 1.5$\angstrom compared to the intrinsic peak positions, and (2) multiple modulated peaks exist within each of the doublet states. Using the underlying hydrodynamical simulation, we could trace back the origin of each subpeak to show how they originate from gas structures or clouds with particular velocity structure.

\subsection{Large-scale Cosmological Volume and Cosmic Web}
\label{sec:cosmo}

\begin{figure*}
    \centering
    \includegraphics[width=0.95\textwidth]{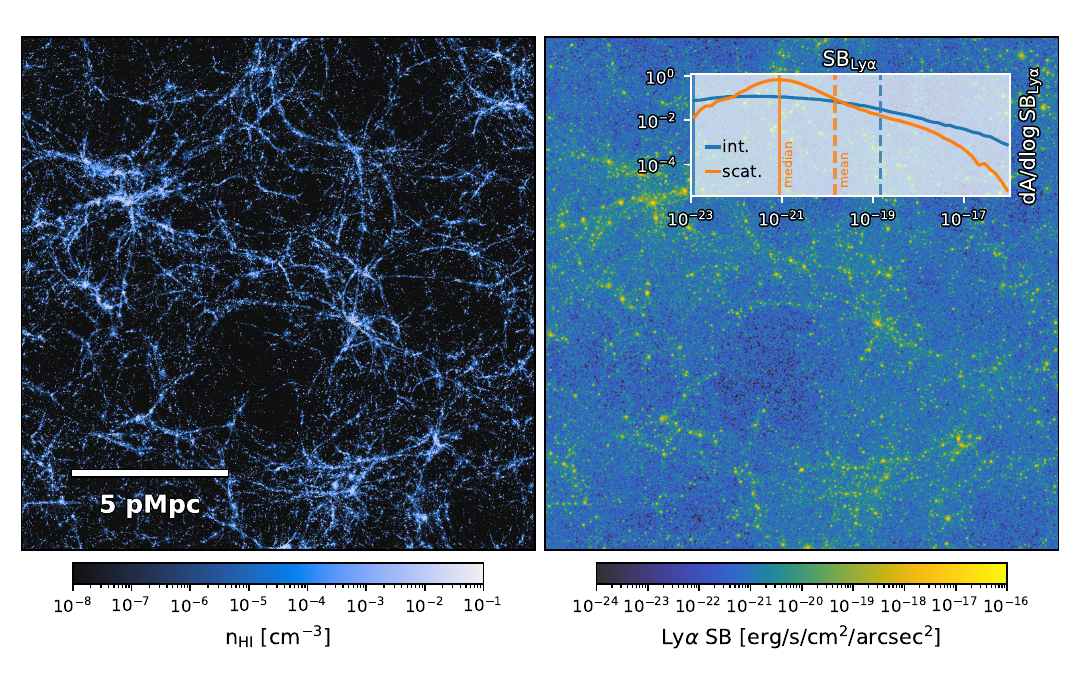} 
    \caption{Application of \thor MCRT to the problem of \Lya emission on the largest scales of the cosmic web. We use the cosmological galaxy formation simulation TNG50 at $z=2$, and apply a simple galaxy-centric \Lya emission model. The true, mean neutral hydrogen density field is shown in projection, using the \thor raytracing operator (left panel). This can be contrasted against the observable, scattered \Lya surface brightness map with a $\sigma=3$\arcsec Gaussian PSF (right panel), where the inset shows the 1-point statistic of the \Lya SB distribution. We show the projection over the total box length. This corresponds to $\Delta\lambda=42$\angstrom in the observed frame, half the \Lya depth of a typical narrow-band survey.}
    \label{fig:cosmosim_large}
\end{figure*}

We now apply \thor to the problem of \Lya radiative transfer across a full, large-volume, cosmological simulation. To do so we turn to TNG50 at $z=2$ \citep{nelson19,pillepich19}. We interpolate the gas fields onto a uniform grid, as before, and adopt the same galaxy-centric, star formation rate based \Lya emission model (Equation \ref{eq:sfrlya}). As we use TNG50-1, this includes galaxies down to $M_\star \gtrsim 10^6$\msun and $\rm{SFR} \gtrsim 10^{-4} \,\rm{M}_\odot \rm{yr}^{-1}$. For each \Lya source we spawn $10^{2}$ photons per $10^{42}$\,erg/s luminosity with a minimum (maximum) photon count of $1$ ($1000$). We impose a global $f_\mathrm{esc}^{\mathrm{Ly}\alpha}=0.1$ for scattered photons, motivated by the lack of explicit dust modeling in this application.

Figure \ref{fig:cosmosim_large} shows large-scale images of the average projected hydrogen density (left) and the scattered \Lya surface brightness (right). The hydrogen density traces the cosmic web, with filamentary structures connecting massive nodes and overdensities. Characteristic values of $n_{\rm HI} > 10^{-1} \rm{cm}^{-3}$ occur within the star-forming regions of galaxies -- our sources -- while extended and low-density gas down to $n_{\rm HI} \sim 10^{-6}$ traces filaments of the cosmic web. These structures are finely resolved in TNG50 and small filaments can be less than $\lesssim 100$pkpc in width.

The \Lya surface brightness traces the same overdense structures and filamentary structures. However, their observable appearance is smoothed out by the numerous scatterings of \Lya photons into the surrounding IGM~\citep{Byrohl23}. This preferentially redistributes \Lya flux from galaxies into their surrounding CGM and IGM. We integrate through the full TNG50 box depth, corresponding to $\sim 42$\angstrom in the observed frame, roughly half of a typical narrow-band filter \citep{ouchi18}. 

In the inset, we show the probability density function of \Lya surface brightness values as the relative area fraction per logarithmic surface brightness dex $dA/d\log\left(\mathrm{SB}_\mathrm{Ly\alpha}\right)$. While the quantitative details depend on spatial smoothing scale as well as spectral width, we find a characteristic value of $\sim 10^{-21}$\sbunits and a rapid decline towards less frequent, higher SB regions. We identify a characteristic behavior where high surface brightnesses from the intrinsic luminosity distribution are shifted towards a more volume filling low-SB regime, in line with previous findings~\citep{Elias20,Byrohl23}. As we only consider emission from compact, star-forming regions, the SB median for the intrinsic emission is zero (placed at $10^{-23}$\sbunits), while the mean is larger than that of the scattered SB distribution. Note that the mean SB changes due to dust attenuation only, which we simplify by the global $f_\mathrm{esc}^{Ly\alpha}$ factor above. 

The observed intensity distribution of \Lya emission on the sky is a key goal of line intensity mapping experiments~\citep{kovetz17,bernal22} and radiative transfer effects play an important role in the case of \Lya in particular~\citep{ambrose25}. For accurate theoretical predictions on large-scale, an empirical calibration step can help constrain unresolved physical processes and degeneracies related to the \Lya sources~\citep[see][for a discussion, as well as \textcolor{blue}{Horaminezhad et al. in prep}]{Byrohl23}.

\subsection{Volume Rendering and Lyman-alpha Forest}

\begin{figure*}
    \includegraphics[width=1.0\textwidth]{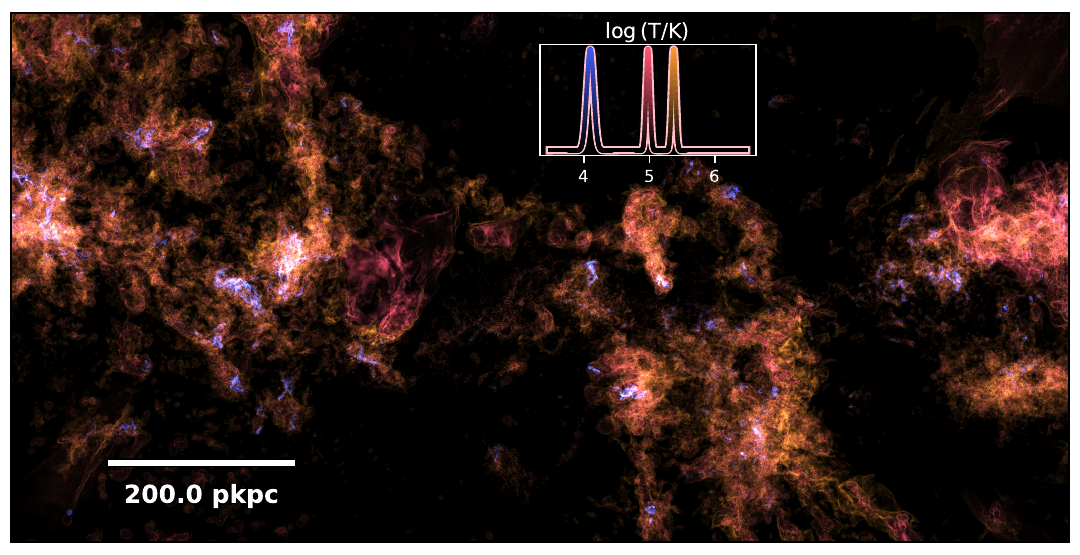}
    \caption{Volume rendering using a simple transfer function for cold (blue) and hot (red/orange) gas in TNG50 centered on the most massive halo at $z=2$. The temperature-dependent transfer function is visualized as a line within the inset where the line color shows the RGB value and the height the alpha value.}
    \label{fig:volumerender}
\end{figure*}

\begin{figure}
    \includegraphics[width=0.46\textwidth]{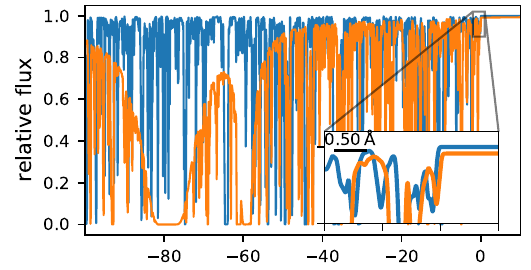}
    \includegraphics[width=0.46\textwidth]{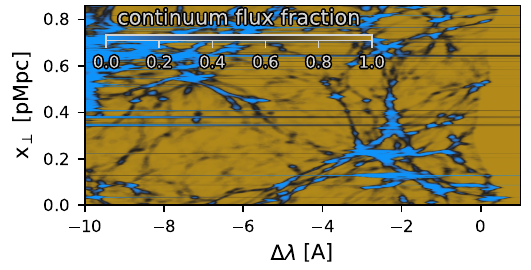}
    \caption{\Lya forest application of \thor to the TNG50 cosmological galaxy formation simulation at $z=2$. \textbf{Top}: Residual continuum flux for two randomly drawn lines of sight using periodic boundary conditions. We show relative flux i.e. with respect to the continuum, as a function of (arbitrary) wavelength. The inset zooms in to highlight spectral detail around the source rest-frame \Lya line-center. Most absorption features are in the weak regime ($\tau_0 \ll 1$). For the second line of sight (orange line), we are in the strong regime ($\tau_0 \gg 1$), traversing a significant hydrogen column densities (N$_\mathrm{HI}\gtrsim 10^{21}$\,cm$^{-2}$). The Lorentzian wings are visible across the full $100$\,\angstrom range. \textbf{Bottom}: \Lya tomography, showing the continuum flux fraction at a given wavelength (x-axis) and a direction perpendicular to the line of sight (y-axis).}
    \label{fig:lyaforest}
\end{figure}

Figure~\ref{fig:volumerender} and~\ref{fig:lyaforest} demonstrate two more advanced operators of the raytracing driver, beyond simple (weighted) projections.

Namely, Figure~\ref{fig:volumerender} shows a volume rendering centered on the most massive halo in the TNG50 simulation at $z=2$. The operator takes a user-defined transfer function, specifying the red, green, blue, and alpha channels (RGBA) as a function of the physical quantity of interest. In this case we define the transfer function as three Gaussian peaks at three discrete values of gas temperature. These highlight cold, warm and hot phases at $\log\left(T\right)=2\cdot 10^4$\,K, $10^5$\,K, and $3\cdot 10^5\,$K in blue, orange, and red, respectively. We also weight the brightness i.e. `emissivity' by gas density, to better highlight structures with substantial mass.

The visualization reveals a protocluster region at high-redshift, with numerous regions of cold gas in and around star-forming galaxies. These systems are surrounding by hotter material due to both feedback and gravitational collapse. In addition to static views, the speed of \thor-based raytracing also enables fast three-dimensional inspection of simulation data, e.g. for scientific exploration and discovery. We can render movies on GPUs with sub-second frame rendering times for $N=1024^3$ data cubes, also enabling quasi-real time interaction.

Figure~\ref{fig:lyaforest} shows a final raytracing operator that produces one-dimensional spectra instead of two-dimensional images. In particular, it calculates mock i.e. synthetic absorption sightlines of neutral hydrogen by raytracing through the simulated distribution of gas in a cosmological volume. Here, we compute an example of a \Lya forest spectra for two random sight lines through the TNG50 simulation box at $z=2$ with periodic boundary conditions. Each gas cell traversed adds a Voigt profile to the spectrum, according to the ground state density of a given species, temperature, and line-of-sight velocity. We sub-sample cells such that each ray step corresponds to a Hubble induced velocity-shift much smaller than the line width (see Section ~\ref{sec:parametersweep}).

We show the resulting spectrum in terms of relative flux as a function of (arbitrary) wavelength. Flux at unity corresponds to total transmission, while a relative flux of zero indicates complete absorption. The characteristic structure of the \Lya forest is visible as a series of narrow absorption features, tracing islands of neutral gas at low-redshift. 
The second line of sight (orange line) intersects two strongly damped systems ($\tau_0\gg 1$), leaving imprints of the Lorentzian wings of the Voigt profile across the spectra. The inset shows the IGM absorption close to the source \Lya line-center, revealing the importance of the IGM on LAE spectra~\citep[also see, e.g.,][]{Byrohl20a}.

The bottom panel shows a two-dimensional `tomographic' view based on many closely spaced sightlines \citep[e.g.][]{Lee14}. Color shows flux, relative to the continuum, as a function of wavelength (x-axis) and one spatial direction (y-axis). Overall, this application demonstrates that the fast, GPU-accelerated capabilities of \thor can be used for many scientific questions beyond resonant line MCRT alone.

\section{Performance and Scaling}
\label{sec:performance}

Having described the design, capabilities, and utility of \thor, we now consider its performance characteristics. We benchmark the code performance, scaling, and efficiency for different target systems that reflect current architectures:

\begin{itemize}
    \item \textbf{HPC}: High-performance computing resources with enterprise CPUs and GPUs. We explore two different test systems. The first has nodes with 2 Intel Xeon Platinum 8360Y at 2.40GHz (72 cores in total) and 4 Nvidia A100 40GB GPUs each. The second has nodes with 2 AMD Instinct MI300A APUs each (48 CPU cores in total).
    \item \textbf{workstation}: Consumer hardware on a desktop-like system. Our test system consists of a AMD Ryzen 5950X at 3.40Ghz (16 cores total), and a Nvidia RTX 3090 24GB.
    \item \textbf{laptop}: Consumer hardware on a laptop-type system. Our test system consists of an Intel i7-1260P (4 performance cores) and an Iris Xe integrated GPU (iGPU).
\end{itemize}

These three broad categories cover most use cases, as well as the major CPU and GPU vendors, integrated versus dedicated GPUs, and consumer versus data-center hardware.

When not stated otherwise, our default configuration is a single socket 36-core Intel Xeon Platinum 8360Y for the CPU backend, and a Nvidia 40GB Ampere A100 as a GPU backend. All tests are run without Simultaneous Multithreading (SMT).\footnote{On the HPC test systems, SMT is deactivated, while on the other systems, we only run at most on $n$ threads equal to the number of CPU cores, and bind these threads to one physical core each.}

We first establish the baseline performance of \thor by comparing to similar, open-source codes in Section~\ref{sec:othercodes}. Second, we compare the CPU and GPU performance in Section~\ref{sec:cpuvsgpu} in order to assess and demonstrate the speed-up possible when using accelerators. Finally, we measure the multi-core and multi-node scaling for distributed work loads in Section~\ref{sec:scaling}.

\subsection{Comparison with other codes}
\label{sec:othercodes}

\begin{table}[ht]
\centering
\caption{Comparison of features between the four MCRT codes we benchmark. We focus on technical capabilities.}
\label{tab:code_features}
\begin{tabular}{lcccc}
\toprule
\textbf{Feature} & \textbf{thor} & \textbf{ILTIS} & \textbf{RASCAS} & \textbf{COLT} \\
\midrule
decomp.\tablefootmark{a} & \cmark & \cmark & \cmark & \xmark \\
multi-data\tablefootmark{b}               & \cmark & \cmark & \xmark & \cmark \\
GPU support\tablefootmark{c}                     & \cmark & \xmark  & \xmark & \xmark \\
higher order\tablefootmark{d}          & \xmark & \xmark & \xmark  & \cmark \\
FP modes\tablefootmark{e}          & \cmark & \xmark & \xmark  & \xmark \\
\midrule
git rev.\tablefootmark{f}  & 3cc51f1 & 9ae77da & aaffc80  & a35fe67 \\
\bottomrule
\end{tabular}
\tablefoot{\tablefoottext{a}{distribute subdomains onto different nodes.}
\tablefoottext{b}{generalized interface to support multiple dataset structures.}
\tablefoottext{c}{optional support for GPU-based computations}
\tablefoottext{d}{reduced floating point precision mode available}
\tablefoottext{e}{higher-order integration to track gradients}
\tablefoottext{f}{git commit hash of reviewed and benchmarked version}
}
\end{table}

We compare our simulations to the several existing and available resonant emission-line codes: COLT, RASCAS, and voroILTIS~\citep{Smith15,Michel-Dansac20,Byrohl21}. Table \ref{tab:code_features} describes their key characteristics, highlighting technical capabilities: support for distributed memory i.e. domain decomposition (first row), support for general inputs beyond those of a specific hydrodynamical code (second row), support for running the MCRT calculation on GPUs (third row), the ability for integrations at higher order than piecewise constant (fourth row), and support for selectable and/or hybrid floating-point precision (last row). \thor is currently unique in its GPU support. The other primary difference between the codes is the data structures that can be used to represent gas fields, e.g. \textsc{Rascas} focuses on AMR-grid geometries, while \textsc{voroIltis} and \textsc{Colt} are the only two methods that currently support unstructured Voronoi tessellations.

\begin{figure}
    \centering
    \includegraphics[width=0.9\columnwidth]{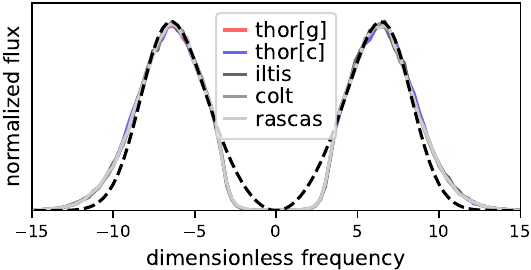}
    \includegraphics[width=0.95\columnwidth]{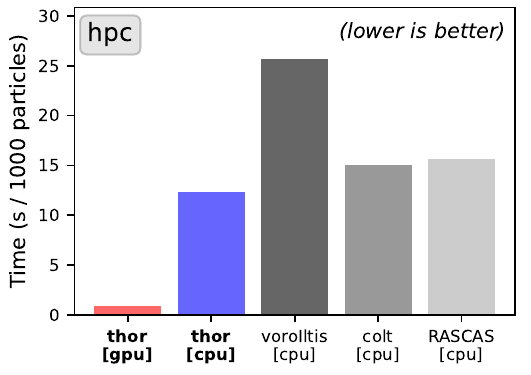}
    \includegraphics[width=0.95\columnwidth]{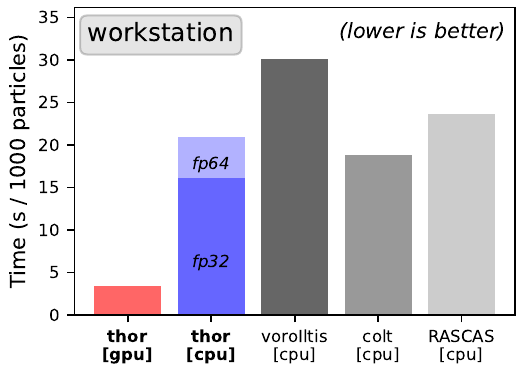}
    \caption{Benchmark and code performance comparison. Here we consider the Neufeld test problem and its analytic solution in the optically thick limit. The top panel shows a comparison of the emergent (final) spectra produced by each code, while the middle (bottom) panels show benchmark results for our HPC (workstation) configurations. For the HPC tests, a single GPU was used, and \thor on the GPU is faster by a factor of $13$. On the workstation, the GPU-to-CPU speedup is $\sim 5$, while FP64 CPU performance is $\sim 20$\% slower than FP32.}
    \label{fig:benchmark_multitarget}
\end{figure}

To compare these codes, we first consider the commonly tested \Lya scattering problem, the \textit{Neufeld} sphere. A sphere of constant neutral hydrogen density is set up at fixed density and temperature. At very high optical depths, an analytic solution exists~\citep{Neufeld90, Dijkstra06}. Here, we fix these parameters to $T=2 \cdot 10^{4}$\,K and a density corresponding to an line-center optical depth of $\tau_0=10^{6}$ along the sphere radius. We always inject $2\cdot 10^5$ photons at line center. 

Figure~\ref{fig:benchmark_multitarget} shows the resulting spectra of escaping photons, comparing the different different MCRT codes (top panel). Overall, they all converge to the same solution. Of the different codes, note that RASCAS is the only case that explicitly realizes the geometry (i.e. creates a grid representing the spherical gas distribution), and this can result in some artifacts. In addition, none of the results exactly reproduce the analytic solution, as the condition $a \tau_0 \gg 1$ is not fully satisfied by the problem setup.

Figure~\ref{fig:benchmark_multitarget} also shows the total runtimes of the five codes, for the HPC setup (middle panel) and workstation setup (bottom panel). When running in CPU mode only, \thor is on par with existing MCRT codes. In fact, all existing codes perform similarly to within a factor of $2$, taking $\sim 10-20$ seconds per 1000 photon packets at this $\tau_0$. However, \thor shows the power of GPU acceleration, demonstrating substantial speed-ups of $\gtrsim 15$ when using the (single A100) GPU backend.\footnote{The speed-up factor primarily relies on the available GPU, as newer GPUs are expected to yield larger speed-ups over their contemporary CPU counterparts. For example, we have run scaling tests on the Nvidia H200 GPU. For this problem, they are $40\%$ faster than the A100, such that a configuration of 4xH200 GPUs and 2x8360Y CPUs will have a GPU-to-CPU speed-up of $\sim 50$.} Consumer GPUs, such as in the lower panel of Figure~\ref{fig:benchmark_multitarget}, usually have significantly less compute power relative to typical CPUs, even at FP32 precision, reducing available speed-ups.

As we will show in Figure~\ref{fig:benchmarks}, the speed-up depends on the column density and acceleration scheme for the \textit{Neufeld} test, as expected.\footnote{We have also tested the code on an integrated GPU (Intel Iris Xe) on a laptop device. While functional, the throughput is significantly lower than on its CPU backend.} As shown, it primarily reflects the $u_\parallel$ implementation of the different codes, and overall performance might significantly vary for more complex scenarios.

For the workstation and laptop consumer targets, we use single-precision floating point precision as double-precision is unavailable (Intel) or has significantly limited hardware support (Nvidia). For the workstation setup, we also show the FP64 CPU performance (light blue), that is roughly $20$\% slower than the FP32 case (dark blue). The workstation GPU \thor run at FP64 is slower than the CPU version by a factor of 3 due to FP64 hardware limitations of the RTX 3090 (not shown).

\subsection{CPU versus GPU}
\label{sec:cpuvsgpu}

We now consider the speed-up of GPU runs over CPU runs with \thor across different setups and performance critical code sections. As a key efficiency indicator, we define the speed-up factor $f=t_\mathrm{CPU}/t_\mathrm{GPU}$ where $t_i$ is the total code walltime, without geometry initialization. In the following, we evaluate and discuss $f$ based on the HPC target (see above) comparing one 36-core Intel CPU to one A100 GPU.\footnote{This choice is somewhat arbitrary: while the nodes we work with have two such CPUs, they have four such GPUs, so our node-level performance would be roughly twice as high GPU-to-CPU speeds-ups compared to what we show below. However, the specifics depend on the system hardware, and we opt for this choice for simplicity.} Note that both CPU and GPU backends run without fast-math compiler flags, and have similar precision.

\begin{figure*}
    \includegraphics[width=0.99\columnwidth]{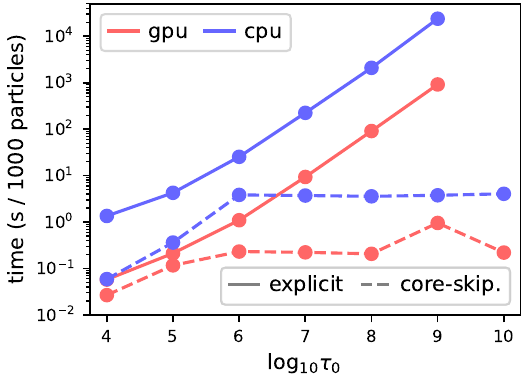}
    \includegraphics[width=0.95\columnwidth]{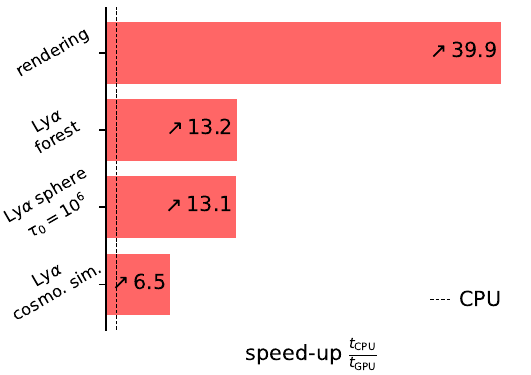}
    \includegraphics[width=1.0\columnwidth]{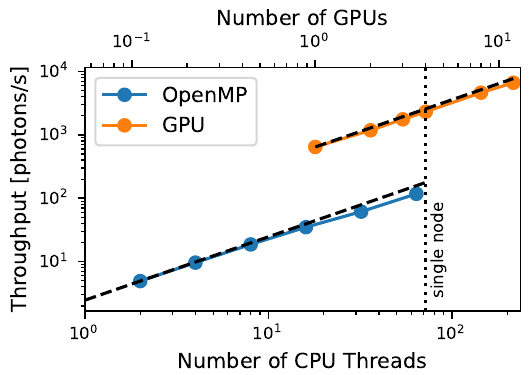}
    \includegraphics[width=1.0\columnwidth]{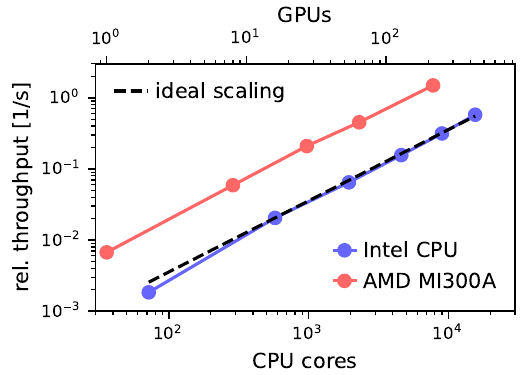}
    \caption{Runtime benchmarks and scaling tests for the \thor radiative transfer code. The application is always \Lya MCRT, although we apply it in several different regimes. \textbf{Upper left:} a spherical setup for varying line-center optical depth $\tau_0$ with and without core-skipping acceleration scheme. We ran $N_\mathrm{photons}=10^4$ ($10^5$) without and $N_\mathrm{photons}=10^5$ ($10^6$) with acceleration scheme on the CPU (GPU) backend. \textbf{Upper right:} the GPU-to-CPU speed-up factor for the Intel CPU + Nvidia GPU HPC system across different problem setups. We find varying speed-ups between a factor 6 and 40 depending on application. \textbf{Lower left:} Strong scaling for a non-distributed setup of the Neufeld $\tau=10^{6}$ problem without core-skipping (see text) and $10^{6}$ ($10^{7}$) photon packages on CPU (GPU). Efficiency for 64 Threads: 74\%. Efficiency for 12 GPUs: 91\%. For the HPC target, there are 36 CPU cores per socket with 2 sockets and 4 GPUs per node. CPU thread scaling decreases as the thread count exceeds a single socket system (efficiency 74\% for 64 threads). We find good scaling of workload to multiple GPUs within and beyond a single node via MPI communication (efficiency 91\% for 12 GPUs). \textbf{Lower right:} Weak scaling test, performed by replicating i.e. tiling the cosmological TNG50 periodic volume starting from a grid size $N_\mathrm{grid}=1024$ (see text). We demonstrate high-efficiency, full-machine scaling, as well as the use of the AMD Instinct accelerators.}
    \label{fig:benchmarks}
\end{figure*}

Figure~\ref{fig:benchmarks} shows a number of quantitative assessments of numerical performance and scaling efficiency.

First, we study the GPU-to-CPU speedup for the shell model as a function of its parameters. The upper left panel of Figure~\ref{fig:benchmarks} shows runtime as a function of the column density $\tau_0$ and whether or not core-skipping is active (upper left panel). As expected, when core-skipping acceleration is disabled, the compute time scales linearly with optical depth at sufficiently high $\tau_0>10^6$. In particular, the number of scatterings is linear in $\tau_0$ for high optical depths. This is true for both CPU and GPU backends. When the core-skipping scheme is activated, the run-time is almost constant. Irrespective of the acceleration scheme, the GPU-to-CPU speedup is $\gtrsim 10$. Below $\tau_0=10^6$, the run-time flattens (steepens) without (with) acceleration. At low column densities and without acceleration, the compute load and kernel times are too small, and GPU performance drops.

It is difficult to anticipate GPU performance for more complex geometries. Figure~\ref{fig:benchmarks} (upper right panel) shows explicit benchmarks of current key \thor applications presented in this paper: resonant emission line MCRT in cosmological simulations and simplified geometries for the \Lya line, absorption spectra for the \Lya forest, and volume rendering. The latter is a classic embarrassingly parallel problem, and we easily achieve speedups of $\sim 40$. In contrast, \Lya MCRT on high dynamic range, multi-scale cosmological simulations is the most challenging case. Typically individual gas cells will have column densities below $\tau_0 \lesssim 10^4$, and the compute work load becomes insufficient for GPUs. However, the propagation logic from cell to cell will ideally keep kernel runtimes sufficiently large and GPU performance high. 

The GPU implementation is optimal when the compute-to-memory-transfer ratio is high and divergence of the compute operations within different threads is low. In \thor MCRT, large divergence occurs due to the frequent sampling of probability distributions via a rejection method. In Appendix~\ref{fig:microbenchmark_uparallel} we investigate the performance of the rejection sampling scheme in its most demanding application: while generating the parallel velocity $u_\parallel$. This divergence is unavoidable, even for identical photon packages in the same location, given their different random sampling pathways. This occurs for a) varying frequencies leading to vastly different regimes in the rejection methods, and b) when the physical environment at the location of photon packages in the same GPU kernel changes. This is particularly true at the start of a Neufeld sphere problem, as all photons start out identical but evolve with time as their frequencies change. Although this can lead to strong divergence, we find in practice that the overall GPU-to-CPU speedups of $\sim 10-20$ remain appreciable.

\subsection{Distributed Scaling}
\label{sec:scaling}

Finally, we consider the scaling performance of \thor as we increase the available compute resources at fixed, and increasing, problem size. Figure~\ref{fig:benchmarks} shows the strong scaling performance for the Neufeld test problem (lower left panel). We consider node-level scaling from 1 to 72 CPU cores (blue), as well as multi-node scaling from 1 to 16 GPUs (orange). The strong scaling performance is only slightly sub-linear in both cases as expected, since the calculation performed by each task is inherently independent. In addition, the single-node GPU-to-CPU speedup is a factor of $\sim 10-20$, showing the significant compute power available with accelerators.

Finally, we demonstrate the distributed scaling of \thor, i.e. the ability to decompose the computational domain onto different machines and GPUs to overcome memory limitations and reduce time to solution. Distributed memory approaches are required to process large hydrodynamical simulations that cannot fit into single device or single node memory. In particular, we perform a weak scaling test of \Lya MCRT using the TNG50 simulation, gridded onto a $1024^3$ data cube, keeping the emission model the same as in Section~\ref{sec:cosmo}. We replicate i.e. tile the volume repeatedly, in order to maintain a constant problem size (workload) per computational node. For example, to test $N_{\rm ranks} = 3^N = 27$ with $N=3$ we replicate the original TNG50-1 volume on each rank forming a contiguous volume $V=27 V_\mathrm{TNG50}$. Photons are propagated until reaching a line-shift of $5$\,\angstrom from the line-center after which we assume the IGM is transparent to \Lya. In order to reach this shift in TNG50 at $z=2$, most \Lya photons need to be communicated to neighboring processes at least once.

Figure~\ref{fig:benchmarks} (lower right) shows the resulting CPU and GPU scaling up to full-machine scale. Our largest jobs use up to 216 machines/GPUs, corresponding to 15k CPU cores. The runs show perfect scaling without any visible bottlenecks due to e.g. memory transfer or communication. Note that in this setup, the work-load is balanced by definition, and this scaling test primarily demonstrates the ability to run large distributed volumes. Performance for $N=1$ is below the ideal scaling due to the lack of domain padding for non-distributed runs.

A strong strength of \thor is its portability and ability to run natively on all common accelerators. This includes not only (Nvidia) GPUs, but also AMD-based accelerated processing units (APUs). In particular, our full-machine scaling tests of Figure~\ref{fig:benchmarks} demonstrate successful, and highly efficient, use of the AMD Instinct MI300A APUs. Here SYCL automatically leverages the HIP backend, and in our benchmarks comparing a single MI300A APU to its single-socket 24 core CPU we obtain a baseline speed-up of $21.3$ times for the Neufeld benchmark.

\section{Discussion} \label{sec:discussion}

\subsection{GPU performance and optimizations}

We have made targeted instruction-level optimizations to improve efficiency, taking into account the variable instruction performance characteristics on different CPU and GPU backends. Here, we highlight key changes when compared to our previous code \textsc{voroILTIS}~\citep{Byrohl21,Byrohl23}.

\textbf{Floating point precision.} \thor can operate at either single or double floating point (FP32, FP64) precision. This allows us to trade precision for moderate speed ups at halved memory requirement. FP32 is particularly useful for applications such as visualization where precision is not a concern. Furthermore, it enables GPU utilization on (consumer) platforms where FP64 evaluation is either much slower, e.g. 16:1 on most Nvidia consumer GPUs, or disabled completely, e.g. integrated Intel GPUs.\footnote{Currently FP32 mode succeeds in all of our regression tests, meaning that consumer GPUs can be used with \thor. This includes \Lya MCRT applications, although caution should be applied in this case.}

\textbf{Branching.} The MCRT algorithm inherently leads to significant code branching, e.g. when a propagation step versus scattering event takes place, as well as during rejection sampling of various probability distributions. This leads to thread divergence~\citep{Cheng14}, stalling execution of work items not following the instruction of the active branch. In Appendix~\ref{sec:uparallel_microbenchmark} we explore the performance of the $u_{\parallel}$ evaluation, our largest source of thread divergence. We counter-intuitively find that GPU backends still largely outperform CPUs, and that pre-sorting photons by frequency improves throughput by a factor of several. 

\textbf{Kernel splitting.} Our MCRT kernel is large and monolithic in design, more so than is common in GPU programming. We have partially split it to separate the initialization of photons and the propagation of peeling contributions. In the future, we expect additional speed ups by dividing out further parts. For example, it might be helpful to propagate photons in homogeneous patches separately from the scattering photon propagation.

\textbf{Instruction choice.} Different hardware configurations for CPUs and GPUs give different capabilities in terms of latency and performance. As such, limitations for non-trivial math operations such as trigonometric functions or roots, can occur on different architectures. While these operations have hardware support on GPUs, the number of concurrent pipelines for some is small, limiting potential performance gains. Similarly, division operations have lower throughput than multiplication on GPUs, and we found it beneficial to replace division and non-trivial math operations whenever possible. While boosting GPU performance in particular, this often improved CPU performance as well. Examples include the choice of approximation for the Voigt profile and the $x_\mathrm{cw}$ computation (see Appendices~\ref{sec:xcw}/\ref{sec:voigt_microbenchmark}).

In addition, instruction-level parallelism (ILP) is an important source of efficiency. For example, when computing higher polynomials other codes commonly leverage Horner's method, which is optimal in terms of math operations, but slower due to inter-operational dependencies. We find that Estrin's method allows for more independent, parallel execution, and can improve performance on both CPUs and GPUs~\citep[see, e.g.,][for a discussion]{Muller16}.

\subsection{Future Directions}

Having demonstrated promising speed-ups for GPU-based MCRT applications in astrophysics, we lay out future directions to address new physics, as well as more advanced numerical algorithms, for applications beyond those we present here.

\subsubsection{Physics}

\thor currently focuses on the physics of gas line emission and the resonant scattering process. Applications include transitions from ultraviolet \Lya~\citep{Byrohl21,Byrohl23} and \MgII~\citep{Nelson21} lines to energetic X-ray lines~\citep[][\textcolor{blue}{Truong et al. in prep}]{Nelson23}. \thor can trivially support other transitions, including doublets and multiplets, in order to calculate multi-wavelength gas emission observables.

\textbf{Emission models.} Natural next steps are to incorporate emission and classically observed optical emission lines from the interstellar medium of galaxies, such as OII and OIII. In addition, the emission from individual stars~\citep[e.g.][]{sanchez06}, stellar populations~\citep[e.g.][]{eldridge17}, dust~\citep[including PAHs, e.g.][]{narayanan23}, and AGN~\citep[e.g.][]{nenkova08} can be included by incorporating available emission models. Notably, AGN can be a significant source of line emission, and the resulting photons can subsequently scatter as they propagate away from the centers of galaxies.\footnote{In addition, the reprocessing of ionizing photons from AGN into recombination radiation from gas could be a significant source of extended emission, e.g. in the case of \Lya \citep{haiman01}. Currently, we capture this effect only if the underlying hydrodynamical simulation includes on-the-fly ionizing RT or has been post-processed by ionizing RT, as \thor is itself not a photoionization code.} We can model AGN by assuming that empirically motivated fractions of their bolometric luminosity are emitted as photons of a particular emission line~\citep[e.g.][]{Cantalupo14,Lusso15,Koptelova17,Costa22}. These are the key steps needed to make quantitative comparisons with high-redshift galaxy observables from e.g. JWST~\citep{stark25}.

\textbf{Molecules and polarization.} At these epochs, ALMA also reveals CO and CII emission as probes of galaxy evolution as well as large-scale intensity mapping, and these are also future targets~\citep[e.g.][]{yue15,inoue20}. In addition, the polarization of \Lya as well as other emission lines is physics easily added to \thor, with various observational applications~\citep{Dijkstra08,hayes11}.

\textbf{Non-static gas state.} At present, when post-processing hydrodynamical simulations, we always treat the underlying fluid fields as static. That is, their thermodynamical state is not updated by radiation. However, many applications of radiative transfer require that we model the impact of radiation on the gas. As the radiation-gas interactions depend on its state, this becomes an iterative problem. Ionizing radiative transfer, NLTE line transfer, and dust radiative transfer are important examples, where equilibrium states have to be solved for during the RT calculation~\citep[e.g.][]{camps15,matsumoto23}.

\textbf{RHD.} This would also pave the way to use \thor as an on-the-fly solver for radiation-hydrodynamical (RHD) simulations~\citep[e.g.][for previous attempts at coupled MCRT-HD simulations]{Vandenbroucke20,smith20}. The impact of small-scale structure in the ISM and CGM~\citep{nelson20,Ramesh24a} on \Lya RT in particular \citep{Hansen06,Gronke17} can be probed and encapsulated via sub-grid RT models (\textcolor{blue}{Byrohl et al. in prep}). 

\subsubsection{Numerics}

The underlying numerical framework of \thor will also benefit from future improvements that will broaden its science applications while increasing performance.

\textbf{Data structures.} We will port our custom geometry library \citep{Byrohl21} to \thor \citep[see also][]{singh25}. This will enable it to natively support data and simulations described with unstructured Voronoi tessellations, as used in many modern astrophysical simulations and datasets~\citep[in particular, AREPO moving-mesh simulations;][]{springel10}. Block and patch-based adaptive mesh refinement (AMR) grids, as well as simple nested Cartesian grids, would likewise enable native support for additional hydrodynamical simulations as well as for problems requiring high spatial dynamic range.

\textbf{Gas state updates.} Many science applications, such as ionization radiative transfer, require concurrent write access to update the fields upon photon traversal. Thread-safe updating requires atomic access that, particularly given the large number of threads available on GPU, can significantly slow down overall execution. Circumventing this bottleneck will require some rework of the underlying dataset access and task scheduling, adopting strategies such as the task-based domain-subdivision presented in~\citep{Vandenbroucke20}. Some of the required infrastructure, such as the use of dependency-managing via task queues, is already available in \thor.

\textbf{Elastic load-balancing.} We currently support MPI-distributed setups for replicated domains (e.g. shell models) as well as distribution of subdomains (e.g. uniform grid) onto different processes/nodes. While load-balancing for replicated domains is available via synchronization of the photon generator state and balancing of photon buffer sizes, this is insufficient for highly inhomogeneous subdomains. Although we can already adjust the subdomain division on initialization based on expected workload, the actual load is hard to predict, especially for resonant MCRT. As a result, adjustment via the creation and removal of subdomains during a run is needed. In its simplest form, this mechanism only needs to replicate an existing domain from disk or via MPI communication. Alternatively, we can dynamically create hotspot-based new subdomains. As we already conceptually differentiate between MPI ranks and the subvolume identifiers, the former case will be straightforward to implement.

\textbf{Compute kernel design.} Nearly all of the computations for the MCRT take place in the same kernel where each work thread computes the photon trajectory over its life cycle. This is a `history-based' approach, as opposed to an `event-based' approach, in which different kernels handle subsequent events, such as uninterrupted propagation or resonant scattering. Previous works in other domains have demonstrated that an event-based approach can (in certain cases) significantly increase performance, as GPU divergence can be largely reduced~\citep{Liu14,Hamilton18,Cuneo24}. Resonant emission line MCRT will require a hybrid approach given its large scattering count, which we will explore in the future. 
Event-based kernels should benefit CPU as well as GPU performance. Currently, \thor and the other tested codes show little x86 AVX2/AVX512-instruction use, as the complex history-based loops are hard to vectorize.
The SYCL kernels can furthermore be adapted to vendor-specific optimizations. Interoperability outside of SYCL kernels needs to be explored to exploit additional raytracing hardware capabilities~\citep{Parker10,Meister24}.

\textbf{Compute kernel preparation.} Major performance improvements can be achieved by preparing data inputs and dynamically adjusting kernel functions. First, data homogeneity within a thread subgroup minimizes divergence. For example, we demonstrate $\gtrsim 2\times$ speedups for pre-sorted drawing of the u$_\parallel$ routine (see Figure~\ref{fig:microbenchmark_uparallel}). Suitable heuristics that sort and group photons by frequency as well as gas state prior to kernel execution are expected to improve performance by even higher margins. Second, we can make use of data-tailored kernels using SYCL's just-in-time compiler, applying optimizations at run-time to recompile suitable kernels. For example, we can choose the most suitable and/or fastest approximation for various algorithmic choices (see Appendix~\ref{sec:optimizations}) given the individual data input. This is feasible as our kernel execution times are much longer than the JIT-compilation process.

\section{Summary} \label{sec:summary}

We introduce the new GPU-accelerated and MPI-parallel Monte Carlo radiative transfer (MCRT) code \thor. Key highlights of its capabilities and performance characteristics are:
\begin{enumerate}
    \item The code runs on heterogeneous hardware architectures and supports all common compute configurations. This is possible via the single-source, multi-target SYCL standard. As a result, \thor can run on CPUs, GPUs and APUs from all common vendors. It can be be used equally on consumer and HPC hardware, and on a laptop, workstation, or large supercomputer cluster.
    \item The code provides abstractions that make it extremely flexible and extensible for various applications. In particular, (1) the driver interface can be customized, and this currently supports MCRT and ray-tracing calculations; (2) the dataset interface specifies input formats and geometries, and we currently support meshless one-dimensional geometries and uniform grids; and (3) the interactor interface defines physical processes related to photons and gas, where our current support includes single and doublet (resonant) emission lines, with and without dust.
    \item Distributed memory calculations are supported via MPI parallelism. We demonstrate applications to up to $6144^{3}$ volume elements and strong and weak scaling up to 100s of GPUs and tens of thousands of CPU cores.
    \item We validate the accuracy and robustness of \thor for resonant emission line radiative transfer with several tests.
    \item We benchmark the performance of \thor on various CPU and GPU configurations. On CPUs, its speed is on par with existing codes. On GPUs, we demonstrate $10-50\times$ speed-ups are possible, depending on the problem and system setup. 
\end{enumerate}

\noindent We then apply \thor to various scientific use cases to showcase the breadth of potential applications. We show radiative transfer calculations for:
\begin{itemize}
    \item Idealized spherical shell models with varying density and velocity profiles, intended for \Lya spectral fitting. We run $\gtrsim 10^{5}$ MCRT calculations across a $9$-dimensional parameter space, using a MCMC sampler to fit a mock spectrum with an asymmetric \Lya double peak.
    \item Post-processing a high-resolution cosmological hydrodynamical zoom simulation of an individual $z=6$ galaxy. The simulation includes a resolved ISM and stellar feedback physics model with $\sim$\,parsec spatial resolution, and we calculate the emergent scattered \Lya surface brightness maps, spectra, and escape fractions.
    \item The circumgalactic medium (CGM) as illuminated by \Lya and \MgII scattering. We post-process a cosmological magnetohydrodynamical (GIBLE) simulation of a Milky Way-like progenitor galaxy at $z \sim 1-2$. We highlight the redistribution of photons due to scattering and the dependence of observables, such as the \Lya spectra and spatially resolved \MgII doublet ratio map, on the viewing direction.
    \item The large-scale cosmic web of gas. We post-processing the TNG50 large-volume cosmological simulation at $z=2$ with \Lya MCRT to capture the scattering of \Lya emission originating from star-forming galaxies. 
    \item GPU-accelerated ray-tracing enables fast projection and volume rendering for visualization. We use the same machinery to calculate synthetic \Lya forest absorption spectra.
\end{itemize}

\noindent Overall, \thor is a modern, highly efficient, and highly versatile radiative transfer code. It forms the foundation for many future applications of MCRT in astrophysical simulations, across a range of scales and physical regimes.

\begin{acknowledgements}
thor has several software dependencies that make it possible. thor is based on C++20 and the SYCL abstraction layer. Primary support is provided via the Clang++ LLVM compiler and the AdaptiveCpp~\citep{Alpay23} SYCL implementation, while also being compatible with the Intel DPCPP implementation. We also make use of the following libraries: CMake, Catch2, z5 (for zarr), HighFive (for HDF5), MPL (for MPI), yamlcpp (for YAML configuration), spdlog/fmt for logging. This work has used the \texttt{scida} analysis library \citep{Byrohl24} for interfacing \thor with different hydrodynamical simulations.

CB thanks Christoph Behrens for many fruitful discussions and previous collaboration on the Iltis radiative transfer code inspiring the development of thor. CB thanks Aaron Smith, Leo Michel-Dansac, and Jeremy Blaizot for discussions on various Lya MCRT implementation aspects. CB thanks Max Gronke for insights on \Lya profile fitting, and Khee-Gan Lee for discussions on \thor's future broader applicability. CB thanks Liam Keegan for discussions on micro-optimization becnhmarking. CB thanks Aksel Alpay and the AdaptiveCpp community for their help on the use of SYCL. We thank Rahul Ramesh for use of the GIBLE simulations. We thank Matthew C. Smith and the MCST team for early use of the high-redshift galaxy formation snapshots. This work was supported by the Deutsche Forschungsgemeinschaft (DFG, German Research Foundation) under Germany's Excellence Strategy EXC 2181/1 - 390900948 (the Heidelberg STRUCTURES Excellence Cluster). CB and DN acknowledge funding from the Deutsche Forschungsgemeinschaft (DFG) through an Emmy Noether Research Group (grant number NE 2441/1-1). CB thanks the Marsilius Kolleg support for financial support for required hardware. CB and DN also thank the Hector Fellow Academy for their funding support. Part of this work was carried out by CB as JSPS International Research Fellow. Calculations were carried out on the Vera, Raven and Viper(GPU) machines of the Max Planck Institute for Astronomy (MPIA) and systems at the Max Planck Computing and Data Facility (MPCDF). The authors acknowledge support by the High Performance and Cloud Computing Group at the Zentrum für Datenverarbeitung of the University of Tübingen, the state of Baden-Württemberg through bwHPC and the German Research Foundation (DFG) through grant no INST 37/1159-1 FUGG.
\end{acknowledgements}

\section*{Data Availability}

Data directly related to this publication is available on request from the corresponding author. The \thor code will be publicly released in the future, and we encourage anyone interested in using, developing, and/or extending \thor to get in touch.

\bibliographystyle{mnras}
\bibliography{references}

\begin{appendix}

\section{Performance critical functions}
\label{sec:optimizations}

In the following, we benchmark and discuss performance critical functions. We provide benchmarks as well as accuracy analysis on CPU and GPU backends. `Micro-benchmarks' of individual functions requires care because of non-trivial optimizations and the execution order of instructions. As a result, synthetic performance tests may not accurately reflect the performance when used in a larger code base. The naive algorithm for micro-benchmarking performance critical functions would be:
\begin{algorithm}
\SetAlgoLined
\KwIn{$n\_outer$ ...}
\KwOut{$data[0\,..\,N{-}1]$}
\For{$i \leftarrow 0$ \KwTo $n\_outer{-}1$}{
    $v \leftarrow \text{compute}(i, ...)$\;
    $data[i] \leftarrow v$\;
}
\caption{Micro-benchmarking}
\label{alg:microbench}
\end{algorithm}
Here, \textsc{compute}(...) is a function to be benchmarked that accepts various arguments, and we perform $n_\mathrm{outer}$ calculations in a loop which are parallelized with a SYCL \textsc{parallel\_for} construct mapping to the respective CPU and GPU backends. Thus, importantly, these benchmarks measure the parallel performance. This is different from commonly used single (CPU) core benchmarking, which however is inapplicable for the GPU backend. Measuring the wall-time of the loop, dividing by the number of loop iterations, gives a measure for the average execution time per computation. 
\begin{algorithm}
\SetAlgoLined
\KwIn{$n\_outer$, $n\_inner$, ...}
\KwOut{$data[0\,..\,N{-}1]$}
\For{$i \leftarrow 0$ \KwTo $N\_outer{-}1$}{
    $v_1 \leftarrow 0$\;
    $k \leftarrow i$\;
    \For{$j \leftarrow 0$ \KwTo $N\_inner - 1$}{
        $v_2 \leftarrow \text{compute}(k, ...)$\;
        \If{$v_2 > v_1$}{
            $v_1 \leftarrow v_2$\;
            $data[i] \leftarrow data[i] + v_1$\;
            $k \leftarrow k + 1$\;
        }
        $k \leftarrow \mathrm{mod}\left(k + 1, N_\mathrm{outer}\right)$\;
    }
}
\caption{Micro-benchmarking with dependency}
\label{alg:microbench_dependency}
\end{algorithm}

However, in the MCRT kernel we are not limited by raw throughput but by the latency each computation call has on subsequent, dependent calculations of a Monte Carlo package. We thus complement above Algorithm~\ref{alg:microbench} by an inner loop with a dependency condition that asserts that a subsequent compute call cannot start before finishing the last (Algorithm~\ref{alg:microbench_dependency}). We compute the average compute time by dividing the wall-time by the product of $n_\mathrm{inner}$ and $n_\mathrm{outer}$.

In short, Algorithm~\ref{alg:microbench} measures throughput, while Algorithm~\ref{alg:microbench_dependency} measures latency. For our MCRT use case, the latency is usually a better performance indicator in the actual application code, and we use this algorithm unless otherwise stated.

\subsection{u$_\parallel$-sampling}
\label{sec:uparallel_microbenchmark}

We test each of the u$_\parallel$ drawing schemes of Section~\ref{sec:driver_mcrt} by generating $10^8$ samples of fixed dimensionless frequency $x$ for an independent random number generator state. We repeat this procedure for different $x\in \left[0.0, 10.0\right]$ and Voigt parameters $a=10^{-2}, 10^{-3}, 10^{-4}$. We run this benchmark for both HPC-CPU and GPU backends. We measure the throughput in iterations per second by dividing the sample count by the kernel execution time. The measurement includes the generation of the thread-local random number generator state but excludes any memory transfer. Note that we use an identical rejection loop for the schemes based on~\citet{Zheng02}. Furthermore, compared to openly available implementations from~\citet{Smith15,Michel-Dansac20}, we optimized the main loop to (i) reduce the number of trigonometric and division operations and (ii) decrease the number of branching operations.

\begin{figure*}
    \includegraphics[width=0.95\textwidth]{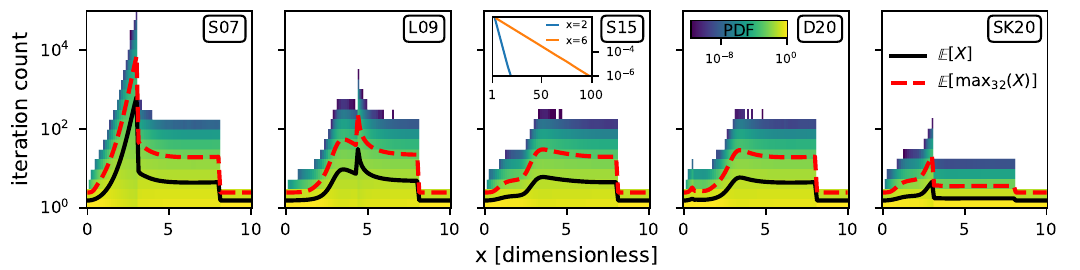}
    \caption{Distribution of rejection iterations needed for different $u_\parallel$ implementations versus dimensionless frequency $x$ for fixed $a=0.001$. The color shows the probability density (per count at fixed $x$). The inset shows the PDF at $x=2$ and $x=6$ for S15, following an exponential distribution. The solid black lines show the expectation value of the iteration count. The dashed red lines show the expectation value of maximum iteration count after drawing 32 times. Given the different hardware architectures, to first approximation, if the iteration-loop dominates the compute time, the inverse of the solid line is proportional to the performance on a CPU, while the inverse of the dashed line is proportional to the performance of a GPU for modern Nvidia, AMD and Intel architectures given their subgroup size of 32.}
    \label{fig:microbenchmark_uparallel_iterationcount}
\end{figure*}

Figure~\ref{fig:microbenchmark_uparallel_iterationcount} shows the number of iterations needed to sample the $u_\parallel$ distribution for different frequencies. We compare five different methods. In all cases, black lines show the expectation value for the iteration count at fixed $x$, while the red dashed lines show the expectation value of the maximum iteration count of 32 draws. While high iteration counts can be needed, across all implementations, these distributions fall exponentially at the high count end, thus substantially limiting the chance for extreme iteration counts. Note that higher iteration counts than shown are possible, but require larger sampling than used ($N=10^{7}$) due to this exponential fall off. As we use rejection sampling, we expect such the PDF to fall off exponentially, given its geometric distribution.

Under the assumption that the function setup was negligible, and no significant instruction level optimizations are present, the black (red) line would be proportional to the CPU (GPU) execution time per draw. The choice of $N=32$ corresponds to the hardware subgroup size for all major GPU vendors (Intel, Nvidia, AMD), and the expectation value of the maximum of the computations within a subgroup reflect the expected divergence by having to wait on the highest iteration count.
We clearly identify the performance bottlenecks for S07 and L09 stemming from the high rejection rates and iteration counts near $x\sim 3$ and $x\sim 4$ respectively.
Across shown $x$-range, we get a mean expectation value ratio of $4.2$ ($3.6$, $3.4$, $3.3$, $2.1$) for S07 (L09, S15, D20, SK20). Hence, we expect divergence causing a significant performance penalty, albeit still more than compensated by the large raw calculation throughput of GPUs.

\begin{figure*}
    \includegraphics[width=0.95\columnwidth]{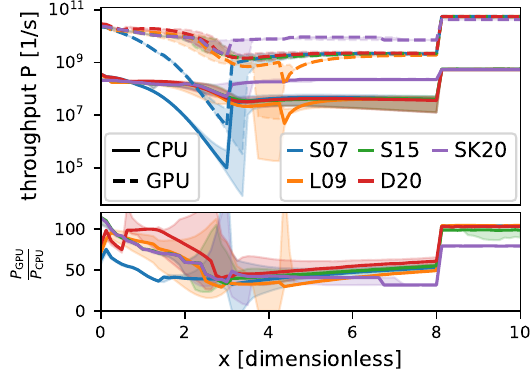}
    \includegraphics[width=0.95\columnwidth]{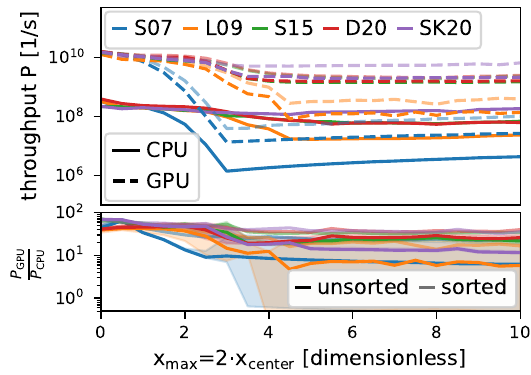}
    \caption{Performance of different implementations for drawing $u_\parallel$ on CPU versus GPU backends. \textbf{Left:} For fixed $x$ and different random seeds. \textbf{Right:} Uniformly randomly sampled $x$ sampled between $0$ and $2\cdot x_\mathrm{center}$. In both cases, the main panels show throughput on one 36-core Intel Xeon Platinum 8360Y CPU and one Nvidia A100 GPU. The bottom sub-panels show the GPU to CPU performance ratio. Different implementations have large variation,  with an average GPU-to-CPU ratio of $\sim 20$. However, large differences in this ratio exist. Lines show the throughput for $a=10^{-3}$, while the shaded band shots the range of outcomes for $10^{-2}$ to $10^{-4}$.}
    \label{fig:microbenchmark_uparallel}
\end{figure*}

Figure~\ref{fig:microbenchmark_uparallel} shows the actual throughput for the respective implementations and backends as a function dimensionless frequency. Shaded bands show the variation in throughput for $a\in\left\{10^{-2}, 10^{-3}, 10^{-4}\right\}$ and lines show the throughput for the $a$ value with median performance. The lower panels provide the throughput ratio of GPU to CPU.

The left panel shows sampling at fixed $x$. In general, performance is best at low $x$ and drops towards intermediate $x\sim 3$, eventually plateauing by $x>8$, where we approximate the distribution by drawing from a Gaussian. S15 and D20 show nearly identical performance on CPU and GPU\footnote{Previous performance differences reported in~\citep{Michel-Dansac20} have been fixed since RASCAS commit \textsc{aaffc80}.}, while the D18 implementation at times behaves slower on the GPU than CPU. SK20 performs significantly better at $x\gtrsim 3$. S07 and L09 show significant performance drops at intermediate $x$. Compared to the CPU backend, the GPU backend shows minimum speedups of $30$, reaching close to $100$ for $x\sim 0$ and $x>8$. The throughput is broadly proportional to the inverse of the iteration count as expected (compare with Figure~\ref{fig:microbenchmark_uparallel_iterationcount}).

Overall, we find strong GPU performance for fixed $x$ and $a$, despite divergent execution paths from different iteration counts. Reaching speedup factors close to $100\times$ is particularly surprising given the expected divergence penalty of $\sim 3\times$.
Note that in realistic MCRT runs, not only $x$ but also $a$ will vary across GPU/CPU threads. 

The right panel of Figure~\ref{fig:microbenchmark_uparallel} compares to the case where the dimensionless frequencies are randomly drawn from a uniform distribution $x\in \left[0.0, 2\cdot x_\mathrm{center}\right]$. Performance tends to degrade towards higher $x$, particularly after reaching $x\gtrsim$ 3. The Single Instruction, Multiple Threads (SIMT) execution model~\citep{Lindholm08} of GPUs penalizes divergent \textit{if} branching and \textit{while} loop counts. We therefore pre-sort the frequencies to reduce thread divergence (faint lines).\footnote{We do not show the pre-sorted throughputs for the CPU backend, which commonly shows a decrease in throughput, most likely related to scheduling imbalances.} The throughput roughly doubles for intermediate $x\sim 3$. S07 and L09 show even larger relative performance gains given their sharp performance drops at some $x$ values. SK20 has a $4\times$ throughput improvement for $x\gtrsim 3$, as we avoid divergence between its largely independent \textit{if} branches. The bottom panel compares pre-sorted GPU to unsorted CPU throughput. Overall, we find that even for varying $x$, the GPU backend yields $\gtrsim 25\times$ higher throughputs, increasing to more than $40\times$ for pre-sorted frequencies.

\subsection{x$_\mathrm{cw}$ estimation}
\label{sec:xcw}

For MCRT we need to estimate the core-to-wing transition from the Gaussian to the Lorentzian of the Voigt profile, for our acceleration scheme as well as $u_\parallel$ generation. This can be explicitly solved for with the Lambert W equation, but a fast approximate implementation for \Lya and a broad temperature range is available~\citep[][S15]{Smith15}. We approximate this solution by a fourth order polynomial (`B25') of $x=\ln(a)$ with $a$ being the Voigt parameter as usual. The polynomial coefficients $c_i$ are $c_0=1.03489162$, $c_1=-6.17680644 \times 10^{-1}$, $c_2=-8.73073343 \times 10^{-2}$, $c_3=-7.91384757 \times 10^{-3}$, $c_4=-2.76037084 \times 10^{-4}$.
Both rely on a logarithm, that we replace with a fast approximation for the floating point format by extracting the exponential bits and approximating the log for the mantissa.\footnote{Inspired by work of Paul Mineiro (github.com/pmineiro/fastapprox).}. Our new scheme `B25fast' is $40$\% ($120$\%) faster on HPC CPU (GPU) backends. Most of this speed-up derives from the log approximation.

\begin{figure}
    \includegraphics[width=0.95\columnwidth]{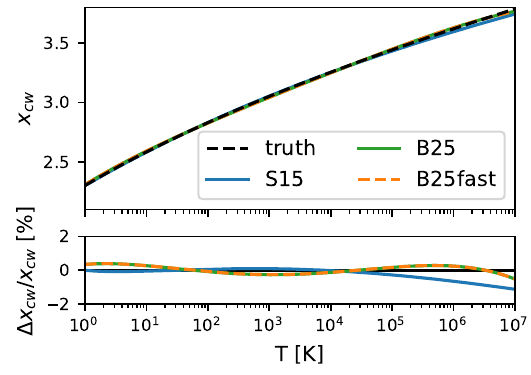}
    \caption{Different approximations for the core-to-wind transition $x_\mathrm{cw}$. We compare S15~\citep{Smith15} with B25 (our method) with and without fastlog approximations (see text). S15 has $<0.1$\% accuracy up to $\sim 10^4\,$K, but this drops to $1$\% at higher temperatures. Our new B25 method has sub-percent accuracy across the temperature range for \Lya, and is substantially faster.}
    \label{fig:xcw}
\end{figure}

Figure~\ref{fig:xcw} compares the precision of different schemes. While S15 on average has better accuracy at $T<10^{5}$\,K, the polynomial approximation is better at higher temperatures with $\lesssim 0.5$\,\% deviation between $1$\,K and $10^7$\,K. 

\begin{figure}
    \includegraphics[width=0.95\columnwidth]{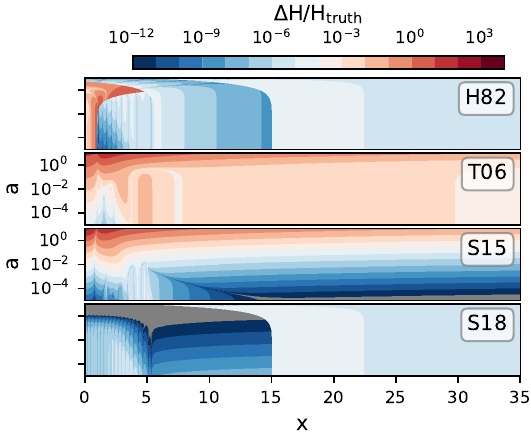}
    \caption{Voigt Line profile error compared to the analytic solution as a function of dimensionless frequency $x$ and the Voigt parameter $a$ for different implementations: H82~\citep{Humlicek82}, T06~\citep{Tasitsiomi06}, S15~\citep{Smith15}, S18~\citep{Schreier18}. We compute the relative error as $\left|H-H_\mathrm{truth}\right|/H_\mathrm{truth}$. Gray errors indicate regions of errors below $<10^{-12}.$}
    \label{fig:lineprofile_err}
\end{figure}

\subsection{Voigt profile}
\label{sec:voigt_microbenchmark}

The Voigt profile is the convolution of a Gaussian and Lorentzian, and has no analytic expression. Different approximations have been introduced for a fast evaluation. The "MIT Faddeeva Package" implementation\footnote{http://ab-initio.mit.edu/faddeeva} provides a close match to the truth with relative errors typically below $10^{-13}$. Although much faster than explicit convolution, the function evaluation is too slow for our frequent evaluation during each propagation step. We therefore consider the following fast approximations: H82~\citep{Humlicek82}, T06~\citep{Tasitsiomi06}, S15~\citep{Smith15}, S18~\citep{Schreier18}. These implementations have different precision and performance characteristics, that vary with dimensionless frequencies and Voigt parameters. We now investigate these properties to guide our default choice in \thor. All implementations are available in \thor at compile time.

Figure~\ref{fig:lineprofile_err} shows the relative error of different numerical implementations of the Voigt profile with the MIT Faddeeva package as the ground truth. Below $a\lesssim 10^{-2}$ (fulfilled for the \Lya line above $\gtrsim 1$\,K) the S15 implementation has an error $\lesssim 10^{-4}$. The H82 and S18 implementations achieve relative errors below $<10^{-4}$ across the full $(x,a)$ parameter space, particularly above $a=10^{-2}$ where S15 accuracy starts to degrade. The commonly used T06 scheme has significantly larger relative errors up to the $1\%$ level for \Lya at $T\sim 1$\,K. H82 and S18 show much better accuracy across the parameter space, particularly at $a\gtrsim 10^{-2}$. While S15 appears sufficient for the \Lya line, emission lines with higher mass and wavelengths, and thus higher Voigt parameters, may require H82 or S18 for sufficient accuracy.

\begin{figure*}
    \includegraphics[width=0.48\textwidth]{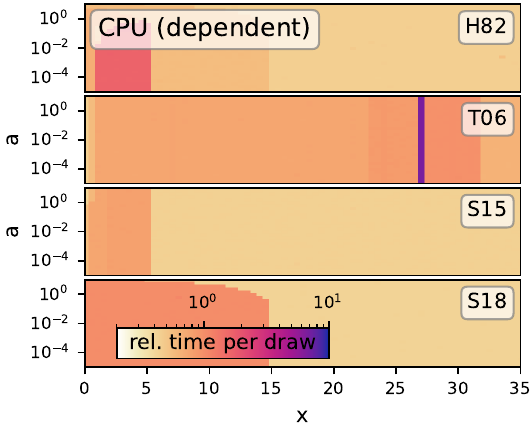}
    \includegraphics[width=0.48\textwidth]{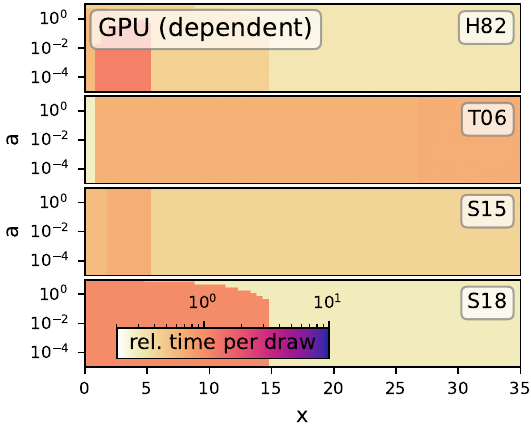}
    \includegraphics[width=0.48\textwidth]{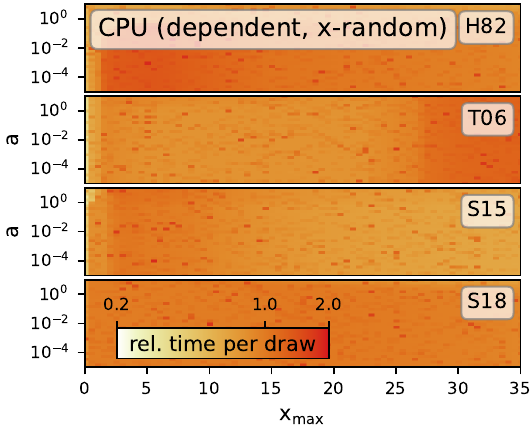}
    \includegraphics[width=0.48\textwidth]{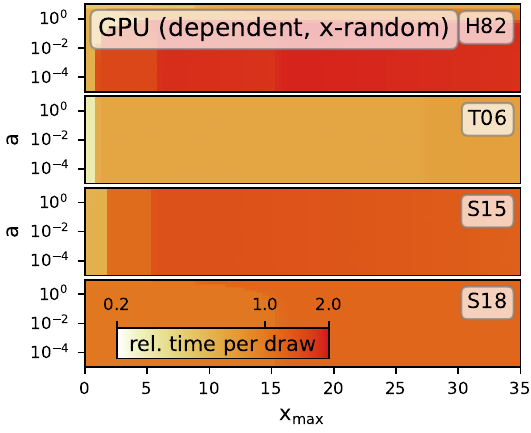}
    \caption{Micro-benchmarking different implementations for the Voigt profile evaluated via Algorithm~\ref{alg:microbench_dependency}. \textbf{Left}: CPU backend (Intel 36 core). \textbf{Right}: GPU backend (Nvidia A100).
    Performance is measured as the time per draw relative to the $x=0, a=0$ timings of S18 on the respective backends. The top row considers draws at fixed dimensionless frequency $x$, while the bottom row considers $x\in\left[0,x_\mathrm{max}\right]$ randomly chosen from a uniform distribution. Distinct regions of relative performance difference correspond to regimes of different approximations and branching within the various implementations.}
    \label{fig:microbenchmark_voigt}
\end{figure*}

Next, Figure~\ref{fig:microbenchmark_voigt} shows the performance of the different implementations when sampling a fixed dimensionless frequency $x$ (top panels). We compare the CPU (left) and GPU (right) backends. We color-code the time per draw relative to the $x=0, a=0$ timings of S18 on each respective backend. The GPU (Nvidia A100) has a speed-up of $\sim 24\times$ over the CPU backend (36 cores).

Importantly, these implementations use different approximations across the $(x,a)$ domain. H82 has if-branched regimes, T06 has two regimes, S15 has three regimes, and S18 has two branches. This broadly aligns with the number of distinct color-coded regions within Figure~\ref{fig:microbenchmark_voigt}. Further substructure is visible for T06 and S15, particularly on the CPU backend where the instruction optimization for some math operations is more often value dependent. This is most notable for T06, showing a tenfold slowdown at $x \simeq 27$ for the CPU backend compared to $(x,a)=(0,0)$, while this is not visible for the GPU backend. We optimize the S18 implementation by rewriting the polynomial evaluation using the Estrin scheme instead of the original Horner evaluation. This yields a speed-up by $110\%$ ($40\%$) on the CPU (GPU) backend. Overall, our CPU results are consistent with previous studies~\citep{Michel-Dansac20}, although differences in relative performance exist, e.g. for T06, potentially due to a custom benchmark algorithm, and different micro-optimizations. Maybe surprisingly, on average, Algorithm~\ref{alg:microbench} performs similar to Algorithm~\ref{alg:microbench_dependency} on the GPU backend, while we find a performance penalty of factor $\sim 2$ on the CPU backend when the dependency is present. \footnote{Note that performance characteristics between the algorithms is more complicated overall. In most cases, the lack of the dependency condition makes timings less pronounced between different if-branches within the same implementation (not shown).} We find a similar behavior for the $u_\parallel$ algorithm in Section~\ref{sec:uparallel_microbenchmark}.

Figure~\ref{fig:microbenchmark_voigt} also shows Voigt profile benchmarks where we instead randomly draw the dimensionless frequency $x\in\left[0,x_\mathrm{max}\right]$ (lower panels). This reflects the more realistic scenario where Monte Carlo contributions take place in the same volume element, but at different frequencies. We again normalize by the $(x,a)=(0,0)$ S18 performance.\footnote{By construction, the speed-up between the GPU and CPU backends for S18 $(0,0)$ is the same as in the top panels.} As expected, performance degrades with increasing $x_\mathrm{max}$, particularly as we encounter additional approximation regimes. This effect is strongest in H82, where we reach an expensive regime at $x\sim 0.9$, and subsequent evaluations for higher $x_\mathrm{max}$ remain costly even after leaving this regime at $x\sim 5.5$.

As GPUs cannot deal efficiently with if-branches, we see a different qualitative behavior in the CPU and GPU panels. Namely, as each branch that is touched upon in a given GPU kernel needs to be effectively executed/waited on by each work element within a thread block, we see increasing evaluation times with every new regime encountered with increasing $x_\mathrm{max}$. We compute the slowdown at $x_\mathrm{max}<0.2$ versus $x_\mathrm{max}>32.0$ at $a<10^{-2}$ by dividing their median execution time. We find a slowdown factor of $1.8$ ($1.6$, $3.5$, $1.2$) for H82 (T06, S15, S18) on the CPU. For the GPU backend, these factors are $3.2$ ($2.5$, $2.3$, $1.2$), respectively. We conclude that there is a noticeable albeit low performance penalty for random $x$ evaluations on GPUs compared to the CPU backend. 

For the MCRT driver, most Voigt profile evaluation occurs near the line-center. We therefore set our fiducial choice based on performance at low $x$. We find T06 to be fastest on both CPU and GPU backend at low $x$, but select S15 at similar performance but with much better accuracy. For applications with many draws away from the line-center (e.g. synthetic absorption spectra), the use of S18 can be beneficial over S15 in terms of performance, in addition its accuracy.

\end{appendix}

\label{lastpage}
\end{document}